\documentclass[aps,amsmath,notitlepage,twocolumn,amssymb,prb,longbibliography]{revtex4-1}

\usepackage[pdftex]{graphicx} 
\usepackage{epstopdf}
\usepackage{verbatim}
\usepackage{color}
\usepackage{subfigure}
\usepackage{tabularx}
\usepackage{amsfonts}
\usepackage{wasysym}
\usepackage{bm}
\usepackage{amsmath}

\usepackage[colorlinks=true,citecolor=red,urlcolor=blue]{hyperref}

\definecolor{darkred}{rgb}{0.847059, 0.141176, 0.164706}
\definecolor{darkgreen}{rgb}{0,0.4,0}
\definecolor{darkblue}{rgb}{0.254902, 0.411765, 0.882353}

\newcolumntype{C}[1]{>{\centering\let\newline\\\arraybackslash\hspace{0pt}}m{#1}}

\begin{document}
\title{The spectral periodicity of the spinon continuum in quantum spin ice}
\author{Gang Chen$^{1,2,3}$}
\email{gangchen.physics@gmail.com}
\affiliation{$^{1}$State Key Laboratory of Surface Physics, Department of Physics,
Fudan University, Shanghai, 200433, China}
\affiliation{$^{2}$Center for Field Theory \& Particle Physics,
Fudan University, Shanghai, 200433, China}
\affiliation{$^{3}$Collaborative Innovation Center of Advanced 
Microstructures, Nanjing, 210093, China}

\date{\today}

\begin{abstract}
Motivated by the rapid experimental progress of quantum spin ice materials, 
we study the dynamical properties of pyrochlore spin ice in the U(1) spin liquid phases. 
In particular, we focus on the spinon excitations that appear at high energies and 
show up as an excitation continuum in the dynamic spin structure factor. 
The keen connection between the crystal symmetry fractionalization of 
the spinons and the spectral periodicity of the spinon continuum is emphasized and 
explicitly demonstrated. When the spinon experiences a background $\pi$ flux 
and the spinon continuum exhibits an enhanced spectral periodicity with a folded Brillouin zone, 
this spectral property can then be used to detect the spin quantum number fractionalization
and U(1) spin liquid. Our prediction can be immediately examined by 
inelastic neutron scattering experiments among quantum spin ice materials with Kramers' 
doublets. Further application to the non-Kramers' doublets is discussed. 
\end{abstract}

\maketitle

\section{Introduction}

The three-dimensional (3D) U(1) quantum spin liquid (QSL) is 
an exotic quantum state of matter and is characterized by fractionalized spinon 
excitation and emergent U(1) gauge structure~\cite{Hermele04}. 
Since the spinons are gapped, the low-energy property of the state 
is described by a compact U(1) quantum electrodynamics in 3D~\cite{Hermele04}. 
This interesting state was proposed more than one decade 
ago~\cite{Hermele04,Motrunich2005,PhysRevLett.91.167004}. 
Recently, there has been a very active search of  
this exotic state among the rare-earth pyrochlore 
quantum spin ice (QSI)~\cite{PhysRevLett.98.157204} materials~\cite{Gingras2014,BalentsSavary,PhysRevLett.105.047201,Savary12,Sungbin2012,SavaryPRB,Gingras2014,Ross2009,Huang2014,PhysRevLett.108.247210,PhysRevB.95.041106,PhysRevB.95.094422,Chen2015,SavaryPRB,PhysRevLett.109.017201,Yasui2002,PhysRevB.64.224416,PhysRevB.90.214430,Chang2012,Kimura2012,RevModPhys.82.53,Lhotel2014,Chang2014,Yasui2003,Ross11,Shannon12,Goswami2016,PhysRevB.95.094407,PhysRevB.94.205107,PhysRevLett.118.107206,Gangchen201706,PhysRevB.92.054432,fu2017fingerprints,PhysRevB.86.075154,PhysRevLett.115.267208,PhysRevLett.109.097205,Frustrating1705,PhysRevLett.107.207207,PhysRevLett.115.097202}. 
Despite the abundance of the QSI materials and possible experimental evidences, 
the identification of U(1) QSL has not been achieved in any candidate material.

To confirm the U(1) QSL, one needs to identify the emergent gauge structure 
and/or the fractionalized spinon excitation. From the theoretical perspective, 
these two things are related since the fractionalized excitation 
naturally emerges in the deconfined phase of the lattice gauge theory. 
Thus, identifying the emergent gauge structure and finding the  
fractionalized spinon excitations are equivalent. For the realistic pyrochlore 
QSIs, the gauge photon and the spinon have drastically different energy 
scales~\cite{Hermele04,Savary12,Sungbin2012}. 
The gauge photon is the very low energy excitation that operates on 
the spin ice manifold~\cite{Shannon12,PhysRevB.86.075154}, 
while the spinons are the much higher energy 
excitations that violate the spin ice rule~\cite{Hermele04}. 
Practically speaking, the large energy-scale difference between 
the gauge photon and spinons suggests that the spinon excitation
might be a better experimental direction to search for. Therefore, 
we focus on the experimental signature of the spinon excitation and 
explore the spectral structure of the spinon continuum in the U(1)  
QSL in this paper. In particular, we point out that the emergent    
background U(1) gauge flux of the ground state enriches the U(1) QSLs 
by creating distinct translational symmetry fractionalization 
for the spinons. In the case that the spinon experiences 
a $\pi$ background flux, there is an enhanced spectral periodicity
with a folded Brillouin zone in the spinon continuum that can be 
revealed by the dynamic spin structure factor in an inelastic 
neutron scattering (INS) measurement. 
The {\it enhanced spectral periodicity} 
is certainly not a property of a conventional paramagnet and thus 
represents {\it an unique experimental signature} of the U(1) QSL 
with the $\pi$ flux.

\begin{figure}[b]
\centering
\includegraphics[width=8.5cm]{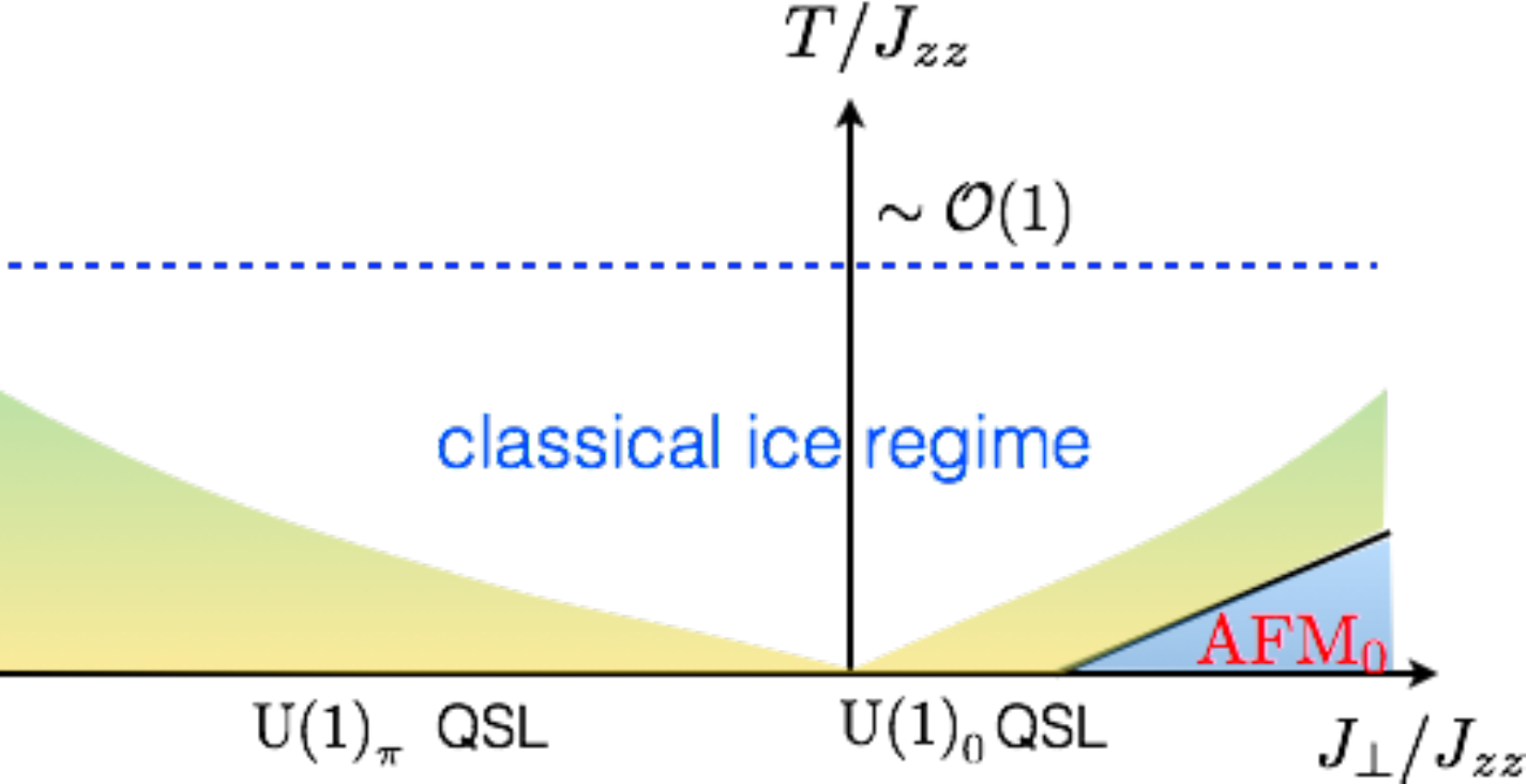}
\caption{(Color online.) The schematic phase diagram of the 
XXZ model on the pyrochlore lattice. The AFM$_0$ stands for 
the magnetic ordered state that is proximate to the ${\text U(1)}_{0}$ 
QSL~\cite{MISC}. The colored region refers to the QSI regime 
in which the quantum fluctuation gradually releases the classical 
spin ice entropy. The QSLs appear as the ground states at zero temperature, 
while the AFM$_0$ extends to finite temperatures.
The solid lines indicate a finite temperature magnetic ordering transition. 
The dashed line indicates the crossover temperature from the high temperature 
paramagnetic regime to the spin ice regime. See the main text and 
Tab.~\ref{tab1} for details. 
}
\label{fig1}
\end{figure}

The enhanced spectral periodicity of the spinon continuum in the U(1) QSL 
with $\pi$ flux, that we discover here, is very analogous to the fractional 
charge excitation in the fractional quantum Hall states~\cite{nature_fqh,Goldman1010}. 
Over there, the global U(1) charge conservation gives the fractional charge 
quantum number to the fractionalized excitation. In our case, it is the 
translation symmetry that is fractionalized and renders  
the enhanced spectral periodicity to the spin continuum. 
Both of these results are examples of symmetry enriched topological orders, 
where the symmetry not only makes the topological ordered phases 
finer~\cite{WenPSG,Wen2002175,PhysRevB.87.104406,PhysRevB.90.121102}
but also makes the topological order more visible from the experimental 
point of view.

The following part of the paper is organized as follows. In Sec.~\ref{sec2},
we introduce the XXZ model as the parent model to extract the $\pi$-flux 
U(1) QSL in the frustrated and perturbative regime. 
In Sec.~\ref{sec3}, we explain the translational symmetry fractionalization and 
predict its consequence on the spectral periodicity of the spinon continuum. 
In Sec.~\ref{sec4}, we explictly compute the spinon continuum with the parton-gauge 
contruction for the XXZ model. 
In Sec.~\ref{sec5}, we discuss the candidate materials and the related experimental
consequences.

\section{Model Hamiltonian and perturbative analysis}
\label{sec2}

\begin{table}[t]
\begin{tabular}{lll}
\hline\hline
U(1) QSLs & $\text{U(1)}_0$ QSL & $\text{U(1)}_{\pi}$ QSL \\ 
\hline
Exchange Coupling & ${J_{\perp}^{} > 0} $ &  ${J_{\perp}^{} < 0}$ \\
Background U(1) Flux & 0 Flux & $\pi$ Flux \\
Heat Capacity & ${C_v \sim T^3}$ & ${C_v \sim T^3}$ \\
Proximate XY Order & Keep Translation  &  Enlarged Cell  \\
Spectral Periodicity & Not Enhanced & Enhanced \\
\hline\hline
\end{tabular}
\caption{Physical properties of the $\text{U(1)}_0$ and $\text{U(1)}_{\pi}$ QSLs.}
\label{tab1}
\end{table}

\begin{figure}[t]
\centering
\includegraphics[width=8.5cm]{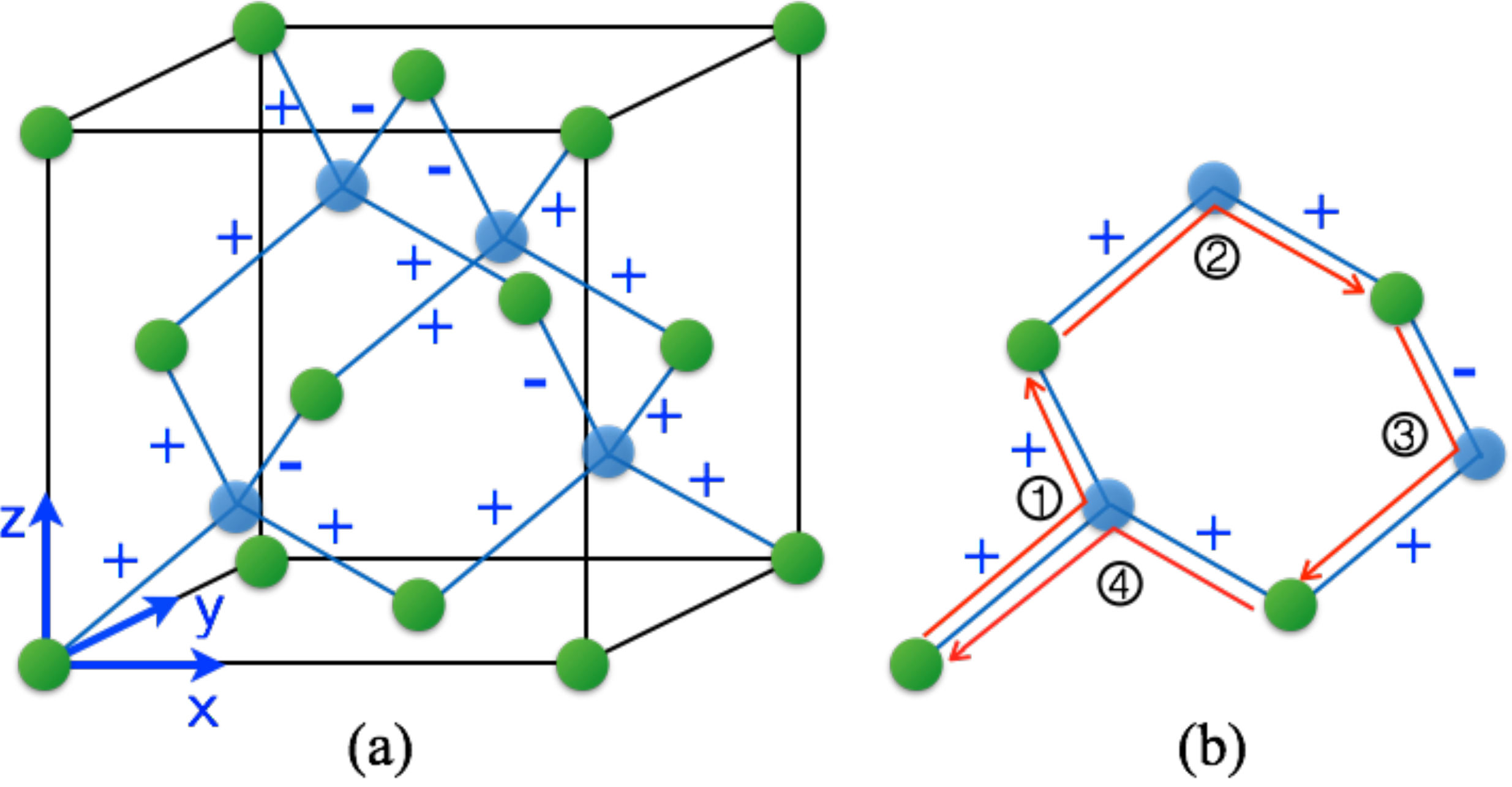}
\caption{(Color online.) The diamond lattice formed 
by the tetrahedral centers of the pyrochlore lattice. 
The dots are the diamond lattice sites or the tetrahedral 
centers of the pyrochlore lattice. 
(a) The spinon hopping for a specific gauge choice 
for the $\pi$ flux. (b) The successive translations 
of the spinon along the (red) pathway, that are marked by 
\textcircled{1}, \textcircled{2}, \textcircled{3} and 
\textcircled{4}, experience the U(1) gauge flux in the 
hexagon plaquette.}
\label{fig2}
\end{figure}

We start with the spin-1/2 XXZ model on the pyrochlore lattice. This model is the parent
model for the pyrochlore QSI~\cite{Hermele04}. The realistic spin models for pyrochlore 
QSI contain more interactions~\cite{Savary12,Sungbin2012,Huang2014,PhysRevB.95.041106}, 
but the simple XXZ model already realizes and captures the generic property 
of the pyrochlore ice U(1) QSL in the perturbative Ising regime. Therefore, we 
deliver our theory through the XXZ model but emphasize the {\it model-independent 
universal and generic} properties of the U(1) QSL. This model is defined as
\begin{eqnarray}
{\mathcal H}_{\text{XXZ}} = \sum_{\langle ij \rangle} 
J_{zz}^{} S^z_i S^z_j - J_{\perp}^{} (S^+_i S^-_j + S^-_i S^+_j),
\label{eq1}
\end{eqnarray}
where ${J_{zz} >0}$. 
The phase diagram of the specific XXZ model is given in Fig.~\ref{fig1}
and explained in the remaining part of the paper. 
In the regime with ${|J_{\perp}| \ll J_{zz}}$, 
the third-order degenerate perturbation theory yields an effective Hamiltonian 
that acts on the extensively degenerate spin ice manifold. 
The effective model is a ring exchange model with~\cite{Hermele04}
\begin{eqnarray}
{\mathcal H}_{\text{eff}} = - \frac{12 J_{\perp}^3}{J_{zz}^2}
\sum_{\hexagon_{\text p}} (S^+_i S^-_j S^+_k S^-_l S^+_m S^-_n + h.c.),
\end{eqnarray}
where ``$i,j,k,l,m,n$'' are the six vertices 
on the elementary hexagon (``$\hexagon_{\text p}$'') 
of the pyrochlore lattice. To reveal the U(1) gauge structure, 
one introduces the lattice gauge fields as 
$E_{{\boldsymbol r}{\boldsymbol r}'} \simeq S^z_{{\boldsymbol r}{\boldsymbol r}'}, 
e^{i A_{{\boldsymbol r}{\boldsymbol r}'}} \simeq 
S^{\pm}_{{\boldsymbol r}{\boldsymbol r}'}$, where 
${\boldsymbol r},{\boldsymbol r}'$ label 
the centers of the tetrahedra
and form a diamond lattice.
The effective spin model becomes
\begin{eqnarray}
{\mathcal H}_{\text{LGT}} = - K 
\sum_{\hexagon_{\text d}} \cos (\text{curl} \, A) 
+ U \sum_{{\boldsymbol r}{\boldsymbol r}'} 
(E_{{\boldsymbol r}{\boldsymbol r}'} 
-\frac{\eta_{\boldsymbol r}}{2})^2, 
\end{eqnarray}
where ${K ={24 J_{\perp}^3}/{J_{zz}^2}}$ 
and ``${U \rightarrow \infty}$'' recovers 
the Hilbert space of the spin-1/2 moment. 
Here ``$\hexagon_{\text d}$'' refers to 
the elementary hexagon on the diamond lattice,
and $\eta_{\boldsymbol r} = + 1$ ($-1$) for 
${{\boldsymbol r}\in}$ I (II) sublattice
of the diamond lattice. When ${J_{\perp} > 0}$
and $|J_{\perp}|$ is small so that the XY order 
is absent, the ground state favors a zero U(1) 
gauge flux and is labeled as U(1)$_{0}$ QSL. This 
regime has been extensively studied theoretically and 
numerically~\cite{Hermele04,Sungbin2012,Savary12,PhysRevLett.100.047208,PhysRevLett.115.037202,PhysRevLett.115.077202,Shannon12}. For ${J_{\perp} < 0}$, the ground state favors
a $\pi$ background U(1) gauge flux with~\cite{Sungbin2012} 
\begin{equation}
\text{curl}\, A\equiv \sum_{{\boldsymbol r}{\boldsymbol r}'
\in \hexagon_{\text d}}{A}_{{\boldsymbol r}{\boldsymbol r}'} 
\!\!\!\!\!\!\!\!\!\!\!\!\!\!\!\!\!\!\!\!\!\!\!\!{\boldsymbol{ \circlearrowleft}}
\quad\quad\quad\,\,
{= \pi}
\end{equation}
for each diamond lattice hexagon (see Fig.~\ref{fig2}a) 
and is thus labeled as U(1)$_{\pi}$ QSL. This regime has 
a sign problem for quantum Monte Carlo simulation and is 
thus less explored. Only one prior work~\cite{Sungbin2012} 
has carefully studied the stability of the U(1) QSL 
in this regime and found the U(1) QSL is more 
robust in this regime than the ${J_{\perp} > 0}$ regime.  
Despite the different phase stability, 
both $\text{U(1)}_{0}$ and $\text{U(1)}_{\pi}$ QSLs are described
by the same low-energy field theory and characterized by the same 
long-distance universal properties. We, however, point out 
that the $\text{U(1)}_{\pi}$ QSL is a distinct symmetry enriched U(1) 
QSL from the $\text{U(1)}_{0}$ QSL. We show below that the symmetry 
enrichment occurs in the translational symmetry fractionalization 
of the spinons. We emphasize that the spectral periodicity of the 
spinon continuum is a keen physical property encoding the distinct symmetry 
enrichment and could thus provide a sharp experimental confirmation 
of the U(1) QSL.

\section{Translational symmetry fractionalization and the spectral periodicity}
\label{sec3}

The translation symmetry of the pyrochlore lattice is generated by the three 
translations $T_1, T_2$, and $T_3$. Here, the $T_{\mu}$ operation translates 
the system by the fcc bravais lattice vector ${\boldsymbol a}_{\mu}$,
and we have ${{\boldsymbol a}_1 = \frac{1}{2}(011)}, 
{{\boldsymbol a}_2 = \frac{1}{2}(101)}$, and ${{\boldsymbol a}_3 = \frac{1}{2}(110)}$, 
where we have used the cubic 
coordinate system here and throughout the paper (except specifically mentioned).   
Any two translation operations, $T_{\mu}$ and $T_{\nu}$ ($\mu \neq \nu$), 
commute with each other with $T_{\mu} T_{\nu} = T_{\nu} T_{\mu}$.

In the U(1) QSL, the spinons are fractionalized and deconfined excitations,
and the symmetry operations act locally on the spinons. This symmetry 
localization condition leads to the symmetry fractionalization for the 
spinons. For the translation symmetry under consideration, we have 
\begin{eqnarray}
T_{\mu}^s T_{\nu}^s = \pm T_{\nu}^s T_{\mu}^s, 
\label{eqcomm}
\end{eqnarray}
where $T_{\mu}^s, T_{\nu}^s$ are the translation operators that operate 
on the individual spinon. The time reversal symmetry demands that 
$T_{\mu}^s$ and $T_{\nu}^s$ either commute or anticommute with each other. 
As the spinon tunnels on the lattice successively following the translation 
operation $T_{\mu}^s T_{\nu}^s (T_{\mu}^s)^{-1} (T_{\nu}^s)^{-1}$, the 
spinon experiences the background U(1) gauge flux. 
If the background U(1) gauge flux is 0 ($\pi$), ``$+$'' (``$-$'') sign is 
chosen in Eq.~\eqref{eqcomm}.

For the XXZ model, it was shown that,
in the regime with ${J_{\perp} < 0}$, each elementary 
hexagon plaquette of the diamond lattice formed by the tetrahedral centers 
traps a $\pi$ U(1) gauge flux for the spinon~\cite{Hermele04}. 
The spinons are created in pairs by the spin flipping operators 
$S^{\pm}$ and reside on the diamond lattice sites of the 
neighboring tetrahedral centers.
It is ready to see from the (red) path on the diamond lattice 
in Fig.~\ref{fig2}b that transporting the spinon according to 
$T_{\mu}^s T_{\nu}^s (T_{\mu}^s)^{-1} (T_{\nu}^s)^{-1}$ experiences 
the same gauge flux in the elementary hexagon plaquette. 
Therefore, for the U(1)$_{\pi}$ QSL with ${J_{\perp} < 0}$, 
we have $T_{\mu}^s T_{\nu}^s = - T_{\nu}^s T_{\mu}^s$. In 
comparison, for the U(1)$_{0}$ QSL with ${J_{\perp} >0}$, we 
have $T_{\mu}^s T_{\nu}^s = + T_{\nu}^s T_{\mu}^s$.

We now explore the sepctroscopic consequence of the non-trivial 
translational symmmetry fractionalization for the spinons in 
the U(1)$_{\pi}$ QSL. To reveal the property of the spinon continuum,
we consider a generic two-spinon scattering 
state~\cite{PhysRevB.90.121102,PhysRevB.65.165113,Wen2002175} 
\begin{equation}
{|a\rangle \equiv |{\boldsymbol q}_{a}; z_a \rangle},
\end{equation}
where ${\boldsymbol q}_{a}$ labels the total crystal momentum
and $z_a$ refers to the remaining quantum numbers such as 
the total energy of the state. Due to the non-orthogonality of the 
fcc bravais lattice vectors, for our convenience we express the momentum 
${\boldsymbol q}_{a}$ as
\begin{equation}
{{\boldsymbol q}_{a} = q_{a1} {\boldsymbol e}_1 + q_{a2} {\boldsymbol e}_2+ q_{a3} {\boldsymbol e}_3}, 
\end{equation}
where ${{\boldsymbol e}_1 = (-1,1,1), 
{\boldsymbol e}_2 = (1,-1,1)}, {{\boldsymbol e}_3 = (1,1,-1)}$,
and $q_{a1}, q_{a2}, q_{a3}$ are the projection of ${\boldsymbol q}_{a}$ 
onto the corresponding ${\boldsymbol e}$ vectors.  
From the symmetry localization condition for the spinons, 
the lattice translation $T_{\mu}$ acts on the state as
\begin{eqnarray}
T_{\mu}^{} |a\rangle = T_{\mu}^s (1) T_{\mu}^s (2) |a\rangle , 
\end{eqnarray} 
where `1' and  `2' label the two spinons, and the translation 
is ``decomposed'' into the two spinon translations. In the following,
we apply the approach that was developed for the 2D $\mathbb{Z}_2$ QSL 
in Ref.~\onlinecite{PhysRevB.90.121102}, but adapt the discussion to our 
3D U(1)$_{\pi}$ QSL. We apply the spinon translation on the
spinon 1 of the state $|a\rangle$ to generate the other
three two-spinon scattering states,
\begin{eqnarray}
|b\rangle &=& T_1^s (1) |a\rangle , \\
|c\rangle &=& T_2^s (1) |a\rangle , \\
|d\rangle &=& T_3^s (1) |a\rangle . 
\end{eqnarray}
All the above states are energy eigenstates and have the same energy
as the two-spinon scattering state $|a\rangle$. 
Nevertheless, these spinon scattering states have distinct crystal momenta. 
To show that, we apply the translation operations on the state $|b\rangle$,
\begin{eqnarray}
T_{1}^{} |b \rangle &=& T_1^s (1) T_1^s (2) T_1^s(1) |a\rangle = +T_1^s (1) [T_1^{} |a\rangle ],  \\
T_{2}^{} |b \rangle &=& T_2^s (1) T_2^s (2) T_1^s(1) |a\rangle = -T_1^s (1) [T_2^{} |a\rangle ], \\
T_{3}^{} |b \rangle &=& T_3^s (1) T_3^s (2) T_1^s(1) |a\rangle = -T_1^s (1) [T_3^{} |a\rangle ], 
\end{eqnarray}
where the anticommutation relation between two spinon translations 
are used in the last two equations. This immediately gives 
\begin{eqnarray}
{q_{b1}^{} = q_{a1}^{}},\quad 
{q_{b2}^{} = {q_{a2}^{} + \pi} }, \quad 
{q_{b3}^{} = {q_{a3}^{} + \pi} }.
\end{eqnarray}
Likewise, we have 
\begin{eqnarray}
q_{c1}^{} &=& {q_{a1}^{} + \pi},\quad 
{q_{c2}^{} = q_{a2}^{} }, \quad 
{q_{c3}^{} = {q_{a3}^{} + \pi}}, \\
q_{d1}^{} &=& {q_{a1}^{} + \pi},\quad 
{q_{d2}^{} = {q_{a2}^{} + \pi}}, \quad 
{q_{d3}^{} = q_{a3}^{}}. 
\end{eqnarray}
The combination of two different spinon translations on $|a\rangle$ such 
as $T_1^s (1)T_2^s (1) |a\rangle$ does not generate new states with 
different momenta. Since the two-spinon scattering states, 
$|a\rangle,|b\rangle,|c\rangle,|d\rangle$, have the same energy and 
the same spin quantum number, the above relations between 
their crystal momenta suggest that, there is an enhanced spectral periodicity 
for the spinon continuum. The spectral periodicity can be reflected by  
the lower ${\mathcal L}({\boldsymbol q})$, upper excitation 
edge ${\mathcal U}({\boldsymbol q})$, and the 
dispersion of the spinon continuum. 
For the U(1)$_{\pi}$ QSL, we have 
\begin{eqnarray}
{\mathcal L}({\boldsymbol q})& =& {\mathcal L}({\boldsymbol q}
+2\pi(100)) = {\mathcal L}({\boldsymbol q}
+2\pi(010)) \nonumber \\ &=& {\mathcal L}({\boldsymbol q}
+2\pi(001)), \\
{\mathcal U}({\boldsymbol q})& =& {\mathcal U}({\boldsymbol q}
+2\pi(100)) = {\mathcal U}({\boldsymbol q}
+2\pi(010)) \nonumber \\ &=& {\mathcal U}({\boldsymbol q}
+2\pi(001)),
\end{eqnarray}
where the momentum and the momentum offset are defined in the 
cubic coordinate system. The spectral intensity of the spinon 
continnum depends on other factors such as the form factor and 
may not respect the enhanced spectral periodicity here.

Usually, the spectral periodicity of the magnetic excitation spectrum is defined 
by the integer mutiples of the reciprocal lattice vectors. Here, because of the 
$\pi$ flux and the translational symmetry fractionalization for the U(1)$_{\pi}$ QSL, 
the Brillouin zone is folded and the spectral periodicity is half of the combination
of two indepedent reciprocal lattice vectors. The spectral periodicity enhancement 
is a rather unique property of the U(1)$_{\pi}$ QSL and is absent in the U(1)$_0$ QSL. 
We emphasize that the enhanced spectral periodicity with a folded Brillouin zone is 
{\it the dynamical property rather than the static property} of the 
U(1)$_{\pi}$ QSL. The U(1)$_{\pi}$ QSL preserves all the lattice 
symmetries, and an elastic neutron scattering would not observe 
any extra magnetic Bragg peak that accompanys with any lattice 
symmetry breaking.

\section{Spinon continuum of the U(1)$_{\pi}$ QSL}
\label{sec4}

Here we 
return to the specific XXZ model and explicitly demonstrate the experimental
consequence of the background $\pi$ flux in the U(1)$_{\pi}$ QSL.
We focus on the ${J_{\perp} < 0}$ regime that has not been extensively studied.    
It was shown that the U(1)$_{\pi}$ QSL extends to 
the point~\cite{Sungbin2012} at ${J_{\perp} = -4.13J_{zz}}$
within a gauge mean-field calculation. 
We, however, do not think the U(1)$_{\pi}$ QSL
can extend beyond the Heisenberg point at ${J_{\perp}=-J_{zz}/2}$ 
where the SU(2) symmetry, that permutes the spin components, is inconsistent with 
the distinct physical meaning of three spin components in the U(1)$_{\pi}$ QSL. 
It is likely that the Heisenberg point is a critical point where 
the U(1)$_{\pi}$ QSL terminates.  Nevertheless, 
the early study does show the quantitative stability of the U(1)$_{\pi}$ QSL. 
Following the previous treatment~\cite{Savary12,Sungbin2012,SavaryPRB,BalentsSavary}, 
we implement the spinon-gauge construction via 
\begin{eqnarray}
S^z_i &=& s^z_{{\boldsymbol r}{\boldsymbol r}'}, \\
S^+_i &=& \Phi^{\dagger}_{\boldsymbol r} \Phi^{\phantom\dagger}_{{\boldsymbol r}'} 
s^+_{{\boldsymbol r}{\boldsymbol r}'} , 
\end{eqnarray}
where $\Phi_{\boldsymbol r}^{\dagger}$ ($\Phi_{\boldsymbol r}^{}$) creates 
(annihilates) the spinon at the diamond lattice site ${\boldsymbol r}$, and 
$s^z$ and $s^{\pm}$ encode the U(1) gauge field 
such that ${s^z_{{\boldsymbol r}{\boldsymbol r}'}
\simeq E_{{\boldsymbol r}{\boldsymbol r}'}^{}}$ 
and ${s^+_{{\boldsymbol r}{\boldsymbol r}'}\simeq \frac{1}{2} 
e^{i A_{{\boldsymbol r}{\boldsymbol r}'}}}$.  
The XXZ model is expressed as 
\begin{eqnarray}
H_{\text{XXZ}} \simeq \frac{J_{zz}}{2} \sum_{\boldsymbol r} Q_{\boldsymbol r}^2 
- \frac{J_{\perp}}{4} \sum_{\langle\langle {\boldsymbol r} {\boldsymbol r}' \rangle\rangle }
\Phi_{{\boldsymbol r}}^{\dagger}
\Phi_{{\boldsymbol r}'}^{}
e^{-iA_{{\boldsymbol r}{\boldsymbol r}'}},
\label{eqham}
\end{eqnarray}
where ${A_{{\boldsymbol r}{\boldsymbol r}'} = A_{{\boldsymbol r}{\boldsymbol r}''}
+ A_{{\boldsymbol r}''{\boldsymbol r}'}}$, and ${\boldsymbol r}''$ is the 
shared nearest neigbhor site of ${\boldsymbol r}$ and ${\boldsymbol r}'$. 
Here the operator $Q_{\boldsymbol r}$ is defined as
${Q_{\boldsymbol r}= \sum_{{\boldsymbol r}'\in n.n.({\boldsymbol r})}
\eta_{\boldsymbol r}^{}  S^z_{{\boldsymbol r}{\boldsymbol r}'}}$,
where the summation is taken for the nearest neighbor sites 
of ${\boldsymbol r}$. A conjugate rotor variable is introduced 
such that 
\begin{equation}
{\Phi_{\boldsymbol r}=e^{-i \phi_{\boldsymbol r}}}, 
\quad 
{|\Phi_{\boldsymbol r}|=1},
\end{equation} 
and ${[\phi_{\boldsymbol r},Q_{\boldsymbol r}] = i}$. 
One further fixes the gauge by setting~\cite{Sungbin2012}
${\bar{A}_{{\boldsymbol r}{\boldsymbol r}'} = {\epsilon_{{\boldsymbol r}{\boldsymbol r}'}
{{\boldsymbol q}_0^{} \cdot {\boldsymbol r}}}}$ 
that takes care of the $\pi$ flux (see Fig.~\ref{fig2}a),
where ${{\boldsymbol q}_0^{} = 2\pi (100)}$, ${\boldsymbol r} \in $ 
I sublattice, and $\epsilon_{{\boldsymbol r}{\boldsymbol r}'}$ 
takes the value 0,1,1,0 for ${\boldsymbol r}{\boldsymbol r}'$
orienting along (111), ($1\bar{1}\bar{1}$), ($\bar{1}1\bar{1}$), ($\bar{1}\bar{1}1$) 
lattice direction, respectively. The gauge fixing condition enlarges 
the unit cell for the spinons, but the translation symmetry is preserved
and is realized projectively. The spinon excitation in 
U(1)$_{\pi}$ QSL is then solved by the standard coherent state
path integral method and is given as~\cite{Sungbin2012}
\begin{eqnarray}
\omega_{\text{I},\pm}^{} ({\boldsymbol k})  &=& \sqrt{2J_{zz} 
\big(\lambda \pm J_{\perp} ({c_y^2 c_z^2 + s_x^2 s_y^2 + c_x^2 s_z^2 })^{\frac{1}{2}} \big) },
\\
\omega_{\text{II},\pm}^{} ({\boldsymbol k}) &=& \sqrt{2J_{zz} 
\big(\lambda \pm J_{\perp} ({s_y^2 s_z^2 + c_x^2 c_y^2 + s_x^2 c_z^2 })^{\frac{1}{2}} \big) },
\end{eqnarray}
where ${c_{\mu} =\cos({k_{\mu}}/{2})}, {s_{\mu} = \sin ({k_{\mu}}/{2})}$. 
The subindices, I, II, arise from the fact that the two diamond lattices 
are decoupled in Eq.~\eqref{eqham} and the subindices, $\pm$, arise
from the doubling of the unit cell by the gauge choice. 
Here, the constraint ${|\Phi_{\boldsymbol r}| = 1}$ is 
implemented by the global Lagrangian multiplier, $\lambda$, 
that is demanded to be uniform for the two 
sublattices by inversion. 

\begin{figure}[t]
\centering
\includegraphics[width=6.4cm]{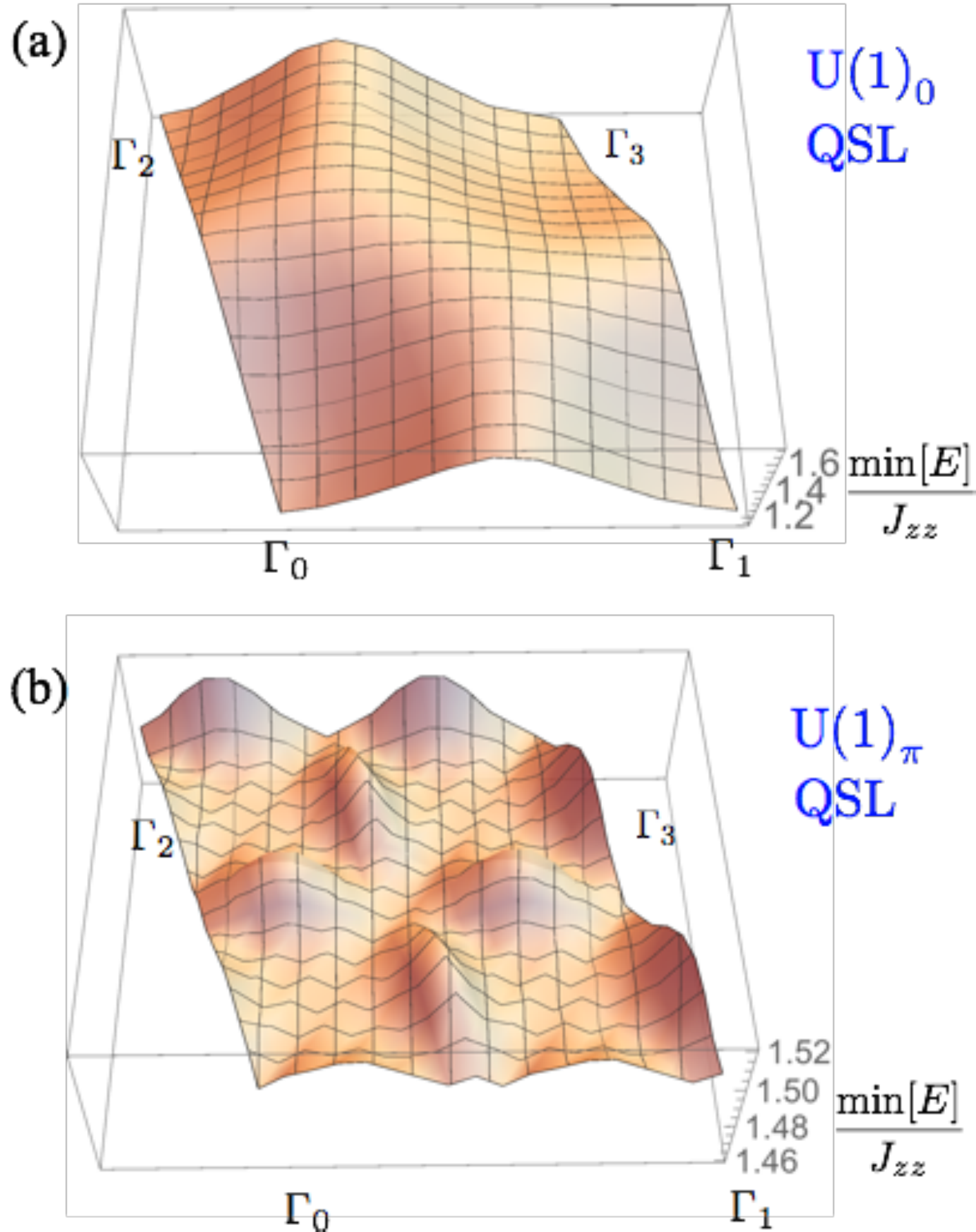}
\caption{(Color online.) The lower excitation edge of the spinon continuum 
in the U(1)$_0$ and the U(1)$_{\pi}$ QSLs. Here, the $\Gamma$ points 
are the centers of the Brillouin zones and are connected by 
the reciprocal lattice vectors with $\Gamma_0\Gamma_1 = 2\pi (-{1},1,1)$ 
and $\Gamma_0\Gamma_2 = 2\pi (1,-{1},1)$. The enhanced spectral periodicity 
in (b) can be visualized by examining the wiggles of the spectrum. 
We set ${J_{\perp} = 0.12J_{zz}}$ for the U(1)$_0$ QSL in (a) and 
${J_{\perp} = -J_{zz}/3}$ for the U(1)$_{\pi}$ QSL in (b).}
\label{fig3}
\end{figure}

The spinon continuum is detected by the $\langle S^+_{i} S^-_j\rangle$ 
correlator via the INS. From the relation 
\begin{eqnarray}
\langle S^+_{i} S^-_j\rangle & \sim & \langle \Phi^\dagger_{{\boldsymbol r}_i} 
\Phi^{}_{{\boldsymbol r}_i'} e^{i A_{ {\boldsymbol r}_i {\boldsymbol r}_i' }}
 \Phi^{}_{{\boldsymbol r}_j} 
\Phi^{\dagger}_{{\boldsymbol r}_j'} e^{-i A_{ {\boldsymbol r}_i {\boldsymbol r}_j' }}
\rangle \nonumber \\
&\simeq& \langle \Phi^\dagger_{{\boldsymbol r}_i}  \Phi^{}_{{\boldsymbol r}_j}  \rangle 
\langle \Phi^{}_{{\boldsymbol r}_i'}  \Phi^{\dagger}_{{\boldsymbol r}_j'} \rangle 
\langle e^{i \bar{A}_{ {\boldsymbol r}_i {\boldsymbol r}_i'} 
          -i \bar{A}_{ {\boldsymbol r}_i {\boldsymbol r}_j'}}\rangle,
\label{eqcorrelation}
\end{eqnarray}
where ${\boldsymbol r}_i, {\boldsymbol r}_j \in$ I,
${\boldsymbol r}_i', {\boldsymbol r}_j' \in$ II, and
the neutron spin flip excites two spinons with one from the I sublattice
and the other from the II sublattice, we obtain the momentum 
and energy transfers of the neutron, 
\begin{eqnarray}
{\boldsymbol q} &=& {\boldsymbol k}_1 + {\boldsymbol k}_2 
                                      + {\boldsymbol q}_0 , \\
{E} &=&  \omega_{{\text I},\mu}^{} ({\boldsymbol k}_1) 
       + \omega_{{\text{II}},\nu}^{} ({\boldsymbol k}_2),
\end{eqnarray}
where ${\mu,\nu =\pm}$ and the offset ${\boldsymbol q}_0$ arises 
from the particular gauge choice for the U(1)$_{\pi}$ QSL, 
and the predicted physical observable does not depend on this choice. 
Here we have neglected the photon contribution that appears 
as a higher order term from the gauge fluctuation with respect 
to the gauge choice in the expansion of Eq.~\eqref{eqcorrelation}. 
The spinons are gapped, and a minimal energy is required to excite them, 
which defines the lower excitation edge. As we show explicitly in 
Fig.~\ref{fig3}, the lower excitation edge of the U(1)$_{\pi}$ QSL has 
the enhanced periodicity while the U(1)$_0$ QSL does not.

\section{Discussion}
\label{sec5}

Although the gapless U(1) gauge photon is one defining feature of the U(1) QSLs, 
its very-low-energy scale and the suppressed spectral weight may prohibit the 
experimental identification~\cite{Savary12,PhysRevB.86.075154}. 
In contrast, the spinon continuum occurs at the higher energy. 
The enhanced spectral periodicity with a fold Brillouin zone of 
the spinon continuum in the U(1)$_{\pi}$ QSL could be 
a sharp signature for the experimental observation. 
Since the U(1)$_{\pi}$ QSL occupies a larger parameter space 
than the U(1)$_0$ QSL\cite{Sungbin2012}, it is thus more likely 
for a candidate system to locate in the U(1)$_{\pi}$ QSL 
and develop the enhanced spectral periodicity that we predict 
for the spinon continuum. 

There are three types of doublets in the rare-earth pyrochlore systems. 
For the non-Kramers doublet like Pr$^{3+}$ in Pr$_2$Ir$_2$O$_7$ and Pr$_2$Zr$_2$O$_7$~\cite{PhysRevB.83.094411,PhysRevLett.105.047201,Sungbin2012,PhysRevB.94.205107} 
since only the Ising component of the local moment is odd under the time reversal, 
the INS would naturally select the Ising components and hence 
only measure the gauge field correlator. The spinon continuum 
cannot be observed for the non-Kramers doublet. For the usual 
Kramers doublet like Yb$^{3+}$ in Yb$_2$Ti$_2$O$_7$~\cite{Ross2009,Ross11,Savary12,SavaryPRB}, 
all the components contribute to the magnetic dipolar moments 
and are thus visible in the INS measurements. Both the gapped spinon 
continuum and the gapless gauge photon are recorded in the INS spectrum. 
As for the dipole-octupole Kramers doublet like Ce$^{3+}$ in 
Ce$_2$Sn$_2$O$_7$~\cite{Huang2014,PhysRevB.95.041106,PhysRevB.94.201114,PhysRevLett.115.097202}, 
it was predicted that~\cite{PhysRevB.95.041106,Huang2014}, 
two distinct symmetry enriched U(1) QSLs, namely, 
the dipolar U(1) QSL and the octupolar U(1) QSL, can occur. 
For the dipolar U(1) QSL, both the gapped spinon 
continuum and the gapless gauge photon show up in the 
INS spectrum, while for the octupolar U(1) QSL, only 
gapped spinon continuum can be detected by the INS~\cite{PhysRevB.95.041106}. 
Therefore, we suggest a careful examination of the spectral 
periodicity of the spinon continuum for the QSI materials 
with Kramers doublets. 

Besides the spectral periodicity, the proximate magnetic order provides 
indirect information about the nearby QSLs. Due to the background $\pi$
flux, we expect the proximate magnetic order of the U(1)$_{\pi}$ QSL 
generically breaks the translation symmetry, while the proximate AFM$_{0}$ 
state of the U(1)$_0$ QSL does not~\cite{Savary12}. 
For the non-Kramers doublets, the spinon condensation from the U(1)$_{\pi}$ 
QSL leads to the transverse spin order that preserves the time reversal 
and corresponds to the magnetic quadrupolar order~\cite{Sungbin2012,PhysRevLett.105.047201}. 
Although the elastic neutron diffraction may not directly probe the quadrupolar 
order and the enlargement of the unit cell, the spin-wave excitation, that 
is created by the dipolar component and detected by INS, reveals the 
intrinsic quadrupolar order and the translation symmetry breaking. 
For the Kramers doublets, the transverse spin order is concomitant 
with the time reversal symmetry breaking and is readily detected by 
neutron diffraction and/or NMR measurements.

\section{Acknowledgements}

I think the anonymous referees for suggestion that improves
the presentation of this paper. I acknowledge Dr.~Jeffrey Rau 
from University of Waterloo for a conversation
and Professor Michel Gingras for a comment. 
I am deeply indebted to Professor Dunghai Lee
whose advice and encouragement motivated me to write out  
and submit my papers including this one. This work is supported 
by the Ministry of Science and Technology of China with the 
Grant No.2016YFA0301001, the Start-Up Funds and the Program 
of First-Class Construction of Fudan University, 
and the Thousand-Youth-Talent Program of China.

\bibliography{refs}

\begin{thebibliography}{54}%
\makeatletter
\providecommand \@ifxundefined [1]{%
 \@ifx{#1\undefined}
}%
\providecommand \@ifnum [1]{%
 \ifnum #1\expandafter \@firstoftwo
 \else \expandafter \@secondoftwo
 \fi
}%
\providecommand \@ifx [1]{%
 \ifx #1\expandafter \@firstoftwo
 \else \expandafter \@secondoftwo
 \fi
}%
\providecommand \natexlab [1]{#1}%
\providecommand \enquote  [1]{``#1''}%
\providecommand \bibnamefont  [1]{#1}%
\providecommand \bibfnamefont [1]{#1}%
\providecommand \citenamefont [1]{#1}%
\providecommand \href@noop [0]{\@secondoftwo}%
\providecommand \href [0]{\begingroup \@sanitize@url \@href}%
\providecommand \@href[1]{\@@startlink{#1}\@@href}%
\providecommand \@@href[1]{\endgroup#1\@@endlink}%
\providecommand \@sanitize@url [0]{\catcode `\\12\catcode `\$12\catcode
  `\&12\catcode `\#12\catcode `\^12\catcode `\_12\catcode `\%12\relax}%
\providecommand \@@startlink[1]{}%
\providecommand \@@endlink[0]{}%
\providecommand \url  [0]{\begingroup\@sanitize@url \@url }%
\providecommand \@url [1]{\endgroup\@href {#1}{\urlprefix }}%
\providecommand \urlprefix  [0]{URL }%
\providecommand \Eprint [0]{\href }%
\providecommand \doibase [0]{http://dx.doi.org/}%
\providecommand \selectlanguage [0]{\@gobble}%
\providecommand \bibinfo  [0]{\@secondoftwo}%
\providecommand \bibfield  [0]{\@secondoftwo}%
\providecommand \translation [1]{[#1]}%
\providecommand \BibitemOpen [0]{}%
\providecommand \bibitemStop [0]{}%
\providecommand \bibitemNoStop [0]{.\EOS\space}%
\providecommand \EOS [0]{\spacefactor3000\relax}%
\providecommand \BibitemShut  [1]{\csname bibitem#1\endcsname}%
\let\auto@bib@innerbib\@empty
\bibitem [{\citenamefont {Hermele}\ \emph {et~al.}(2004)\citenamefont
  {Hermele}, \citenamefont {Fisher},\ and\ \citenamefont
  {Balents}}]{Hermele04}%
  \BibitemOpen
  \bibfield  {author} {\bibinfo {author} {\bibfnamefont {Michael}\ \bibnamefont
  {Hermele}}, \bibinfo {author} {\bibfnamefont {Matthew P.~A.}\ \bibnamefont
  {Fisher}}, \ and\ \bibinfo {author} {\bibfnamefont {Leon}\ \bibnamefont
  {Balents}},\ }\bibfield  {title} {\enquote {\bibinfo {title} {{Pyrochlore
  photons: The $U(1)$ spin liquid in a $S=\frac{1}{2}$ three-dimensional
  frustrated magnet}},}\ }\href {\doibase 10.1103/PhysRevB.69.064404}
  {\bibfield  {journal} {\bibinfo  {journal} {Phys. Rev. B}\ }\textbf {\bibinfo
  {volume} {69}},\ \bibinfo {pages} {064404} (\bibinfo {year}
  {2004})}\BibitemShut {NoStop}%
\bibitem [{\citenamefont {Motrunich}\ and\ \citenamefont
  {Senthil}(2005)}]{Motrunich2005}%
  \BibitemOpen
  \bibfield  {author} {\bibinfo {author} {\bibfnamefont {O.~I.}\ \bibnamefont
  {Motrunich}}\ and\ \bibinfo {author} {\bibfnamefont {T.}~\bibnamefont
  {Senthil}},\ }\bibfield  {title} {\enquote {\bibinfo {title} {{Origin of
  artificial electrodynamics in three-dimensional bosonic models}},}\ }\href
  {\doibase 10.1103/PhysRevB.71.125102} {\bibfield  {journal} {\bibinfo
  {journal} {Phys. Rev. B}\ }\textbf {\bibinfo {volume} {71}},\ \bibinfo
  {pages} {125102} (\bibinfo {year} {2005})}\BibitemShut {NoStop}%
\bibitem [{\citenamefont {Huse}\ \emph {et~al.}(2003)\citenamefont {Huse},
  \citenamefont {Krauth}, \citenamefont {Moessner},\ and\ \citenamefont
  {Sondhi}}]{PhysRevLett.91.167004}%
  \BibitemOpen
  \bibfield  {author} {\bibinfo {author} {\bibfnamefont {David~A.}\
  \bibnamefont {Huse}}, \bibinfo {author} {\bibfnamefont {Werner}\ \bibnamefont
  {Krauth}}, \bibinfo {author} {\bibfnamefont {R.}~\bibnamefont {Moessner}}, \
  and\ \bibinfo {author} {\bibfnamefont {S.~L.}\ \bibnamefont {Sondhi}},\
  }\bibfield  {title} {\enquote {\bibinfo {title} {{Coulomb and Liquid Dimer
  Models in Three Dimensions}},}\ }\href {\doibase
  10.1103/PhysRevLett.91.167004} {\bibfield  {journal} {\bibinfo  {journal}
  {Phys. Rev. Lett.}\ }\textbf {\bibinfo {volume} {91}},\ \bibinfo {pages}
  {167004} (\bibinfo {year} {2003})}\BibitemShut {NoStop}%
\bibitem [{\citenamefont {Molavian}\ \emph {et~al.}(2007)\citenamefont
  {Molavian}, \citenamefont {Gingras},\ and\ \citenamefont
  {Canals}}]{PhysRevLett.98.157204}%
  \BibitemOpen
  \bibfield  {author} {\bibinfo {author} {\bibfnamefont {Hamid~R.}\
  \bibnamefont {Molavian}}, \bibinfo {author} {\bibfnamefont {Michel J.~P.}\
  \bibnamefont {Gingras}}, \ and\ \bibinfo {author} {\bibfnamefont {Benjamin}\
  \bibnamefont {Canals}},\ }\bibfield  {title} {\enquote {\bibinfo {title}
  {{Dynamically Induced Frustration as a Route to a Quantum Spin Ice State in
  ${\mathrm{Tb}}_{2}{\mathrm{Ti}}_{2}{\mathrm{O}}_{7}$ via Virtual Crystal
  Field Excitations and Quantum Many-Body Effects}},}\ }\href {\doibase
  10.1103/PhysRevLett.98.157204} {\bibfield  {journal} {\bibinfo  {journal}
  {Phys. Rev. Lett.}\ }\textbf {\bibinfo {volume} {98}},\ \bibinfo {pages}
  {157204} (\bibinfo {year} {2007})}\BibitemShut {NoStop}%
\bibitem [{\citenamefont {Gingras}\ and\ \citenamefont
  {McClarty}(2014)}]{Gingras2014}%
  \BibitemOpen
  \bibfield  {author} {\bibinfo {author} {\bibfnamefont {M~J~P}\ \bibnamefont
  {Gingras}}\ and\ \bibinfo {author} {\bibfnamefont {P~A}\ \bibnamefont
  {McClarty}},\ }\bibfield  {title} {\enquote {\bibinfo {title} {{Quantum spin
  ice: a search for gapless quantum spin liquids in pyrochlore magnets}},}\
  }\href {http://stacks.iop.org/0034-4885/77/i=5/a=056501} {\bibfield
  {journal} {\bibinfo  {journal} {Reports on Progress in Physics}\ }\textbf
  {\bibinfo {volume} {77}},\ \bibinfo {pages} {056501} (\bibinfo {year}
  {2014})}\BibitemShut {NoStop}%
\bibitem [{\citenamefont {Savary}\ and\ \citenamefont
  {Balents}(2016)}]{BalentsSavary}%
  \BibitemOpen
  \bibfield  {author} {\bibinfo {author} {\bibfnamefont {Lucile}\ \bibnamefont
  {Savary}}\ and\ \bibinfo {author} {\bibfnamefont {Leon}\ \bibnamefont
  {Balents}},\ }\bibfield  {title} {\enquote {\bibinfo {title} {{Quantum spin
  liquids: a review}},}\ }\href@noop {} {\bibfield  {journal} {\bibinfo
  {journal} {Reports on Progress in Physics}\ }\textbf {\bibinfo {volume}
  {80}},\ \bibinfo {pages} {016502} (\bibinfo {year} {2016})}\BibitemShut
  {NoStop}%
\bibitem [{\citenamefont {Onoda}\ and\ \citenamefont
  {Tanaka}(2010)}]{PhysRevLett.105.047201}%
  \BibitemOpen
  \bibfield  {author} {\bibinfo {author} {\bibfnamefont {Shigeki}\ \bibnamefont
  {Onoda}}\ and\ \bibinfo {author} {\bibfnamefont {Yoichi}\ \bibnamefont
  {Tanaka}},\ }\bibfield  {title} {\enquote {\bibinfo {title} {{Quantum Melting
  of Spin Ice: Emergent Cooperative Quadrupole and Chirality}},}\ }\href
  {\doibase 10.1103/PhysRevLett.105.047201} {\bibfield  {journal} {\bibinfo
  {journal} {Phys. Rev. Lett.}\ }\textbf {\bibinfo {volume} {105}},\ \bibinfo
  {pages} {047201} (\bibinfo {year} {2010})}\BibitemShut {NoStop}%
\bibitem [{\citenamefont {Savary}\ and\ \citenamefont
  {Balents}(2012)}]{Savary12}%
  \BibitemOpen
  \bibfield  {author} {\bibinfo {author} {\bibfnamefont {Lucile}\ \bibnamefont
  {Savary}}\ and\ \bibinfo {author} {\bibfnamefont {Leon}\ \bibnamefont
  {Balents}},\ }\bibfield  {title} {\enquote {\bibinfo {title} {{Coulombic
  Quantum Liquids in Spin-$1/2$ Pyrochlores}},}\ }\href {\doibase
  10.1103/PhysRevLett.108.037202} {\bibfield  {journal} {\bibinfo  {journal}
  {Phys. Rev. Lett.}\ }\textbf {\bibinfo {volume} {108}},\ \bibinfo {pages}
  {037202} (\bibinfo {year} {2012})}\BibitemShut {NoStop}%
\bibitem [{\citenamefont {Lee}\ \emph {et~al.}(2012)\citenamefont {Lee},
  \citenamefont {Onoda},\ and\ \citenamefont {Balents}}]{Sungbin2012}%
  \BibitemOpen
  \bibfield  {author} {\bibinfo {author} {\bibfnamefont {SungBin}\ \bibnamefont
  {Lee}}, \bibinfo {author} {\bibfnamefont {Shigeki}\ \bibnamefont {Onoda}}, \
  and\ \bibinfo {author} {\bibfnamefont {Leon}\ \bibnamefont {Balents}},\
  }\bibfield  {title} {\enquote {\bibinfo {title} {{Generic quantum spin
  ice}},}\ }\href {\doibase 10.1103/PhysRevB.86.104412} {\bibfield  {journal}
  {\bibinfo  {journal} {Phys. Rev. B}\ }\textbf {\bibinfo {volume} {86}},\
  \bibinfo {pages} {104412} (\bibinfo {year} {2012})}\BibitemShut {NoStop}%
\bibitem [{\citenamefont {Savary}\ and\ \citenamefont
  {Balents}(2013)}]{SavaryPRB}%
  \BibitemOpen
  \bibfield  {author} {\bibinfo {author} {\bibfnamefont {Lucile}\ \bibnamefont
  {Savary}}\ and\ \bibinfo {author} {\bibfnamefont {Leon}\ \bibnamefont
  {Balents}},\ }\bibfield  {title} {\enquote {\bibinfo {title} {{Spin liquid
  regimes at nonzero temperature in quantum spin ice}},}\ }\href {\doibase
  10.1103/PhysRevB.87.205130} {\bibfield  {journal} {\bibinfo  {journal} {Phys.
  Rev. B}\ }\textbf {\bibinfo {volume} {87}},\ \bibinfo {pages} {205130}
  (\bibinfo {year} {2013})}\BibitemShut {NoStop}%
\bibitem [{\citenamefont {Ross}\ \emph {et~al.}(2009)\citenamefont {Ross},
  \citenamefont {Ruff}, \citenamefont {Adams}, \citenamefont {Gardner},
  \citenamefont {Dabkowska}, \citenamefont {Qiu}, \citenamefont {Copley},\ and\
  \citenamefont {Gaulin}}]{Ross2009}%
  \BibitemOpen
  \bibfield  {author} {\bibinfo {author} {\bibfnamefont {K.~A.}\ \bibnamefont
  {Ross}}, \bibinfo {author} {\bibfnamefont {J.~P.~C.}\ \bibnamefont {Ruff}},
  \bibinfo {author} {\bibfnamefont {C.~P.}\ \bibnamefont {Adams}}, \bibinfo
  {author} {\bibfnamefont {J.~S.}\ \bibnamefont {Gardner}}, \bibinfo {author}
  {\bibfnamefont {H.~A.}\ \bibnamefont {Dabkowska}}, \bibinfo {author}
  {\bibfnamefont {Y.}~\bibnamefont {Qiu}}, \bibinfo {author} {\bibfnamefont
  {J.~R.~D.}\ \bibnamefont {Copley}}, \ and\ \bibinfo {author} {\bibfnamefont
  {B.~D.}\ \bibnamefont {Gaulin}},\ }\bibfield  {title} {\enquote {\bibinfo
  {title} {{Two-Dimensional Kagome Correlations and Field Induced Order in the
  Ferromagnetic $XY$ Pyrochlore
  ${\mathrm{Yb}}_{2}{\mathrm{Ti}}_{2}{\mathrm{O}}_{7}$}},}\ }\href {\doibase
  10.1103/PhysRevLett.103.227202} {\bibfield  {journal} {\bibinfo  {journal}
  {Phys. Rev. Lett.}\ }\textbf {\bibinfo {volume} {103}},\ \bibinfo {pages}
  {227202} (\bibinfo {year} {2009})}\BibitemShut {NoStop}%
\bibitem [{\citenamefont {Huang}\ \emph {et~al.}(2014)\citenamefont {Huang},
  \citenamefont {Chen},\ and\ \citenamefont {Hermele}}]{Huang2014}%
  \BibitemOpen
  \bibfield  {author} {\bibinfo {author} {\bibfnamefont {Yi-Ping}\ \bibnamefont
  {Huang}}, \bibinfo {author} {\bibfnamefont {Gang}\ \bibnamefont {Chen}}, \
  and\ \bibinfo {author} {\bibfnamefont {Michael}\ \bibnamefont {Hermele}},\
  }\bibfield  {title} {\enquote {\bibinfo {title} {{Quantum Spin Ices and
  Topological Phases from Dipolar-Octupolar Doublets on the Pyrochlore
  Lattice}},}\ }\href {\doibase 10.1103/PhysRevLett.112.167203} {\bibfield
  {journal} {\bibinfo  {journal} {Phys. Rev. Lett.}\ }\textbf {\bibinfo
  {volume} {112}},\ \bibinfo {pages} {167203} (\bibinfo {year}
  {2014})}\BibitemShut {NoStop}%
\bibitem [{\citenamefont {Wan}\ and\ \citenamefont
  {Tchernyshyov}(2012)}]{PhysRevLett.108.247210}%
  \BibitemOpen
  \bibfield  {author} {\bibinfo {author} {\bibfnamefont {Yuan}\ \bibnamefont
  {Wan}}\ and\ \bibinfo {author} {\bibfnamefont {Oleg}\ \bibnamefont
  {Tchernyshyov}},\ }\bibfield  {title} {\enquote {\bibinfo {title} {{Quantum
  Strings in Quantum Spin Ice}},}\ }\href {\doibase
  10.1103/PhysRevLett.108.247210} {\bibfield  {journal} {\bibinfo  {journal}
  {Phys. Rev. Lett.}\ }\textbf {\bibinfo {volume} {108}},\ \bibinfo {pages}
  {247210} (\bibinfo {year} {2012})}\BibitemShut {NoStop}%
\bibitem [{\citenamefont {Li}\ and\ \citenamefont
  {Chen}(2017)}]{PhysRevB.95.041106}%
  \BibitemOpen
  \bibfield  {author} {\bibinfo {author} {\bibfnamefont {Yao-Dong}\
  \bibnamefont {Li}}\ and\ \bibinfo {author} {\bibfnamefont {Gang}\
  \bibnamefont {Chen}},\ }\bibfield  {title} {\enquote {\bibinfo {title}
  {{Symmetry enriched $U(1)$ topological orders for dipole-octupole doublets on
  a pyrochlore lattice}},}\ }\href {\doibase 10.1103/PhysRevB.95.041106}
  {\bibfield  {journal} {\bibinfo  {journal} {Phys. Rev. B}\ }\textbf {\bibinfo
  {volume} {95}},\ \bibinfo {pages} {041106} (\bibinfo {year}
  {2017})}\BibitemShut {NoStop}%
\bibitem [{\citenamefont {Yan}\ \emph {et~al.}(2017)\citenamefont {Yan},
  \citenamefont {Benton}, \citenamefont {Jaubert},\ and\ \citenamefont
  {Shannon}}]{PhysRevB.95.094422}%
  \BibitemOpen
  \bibfield  {author} {\bibinfo {author} {\bibfnamefont {Han}\ \bibnamefont
  {Yan}}, \bibinfo {author} {\bibfnamefont {Owen}\ \bibnamefont {Benton}},
  \bibinfo {author} {\bibfnamefont {Ludovic}\ \bibnamefont {Jaubert}}, \ and\
  \bibinfo {author} {\bibfnamefont {Nic}\ \bibnamefont {Shannon}},\ }\bibfield
  {title} {\enquote {\bibinfo {title} {{Theory of multiple-phase competition in
  pyrochlore magnets with anisotropic exchange with application to
  ${\mathrm{Yb}}_{2}{\mathrm{Ti}}_{2}{\mathrm{O}}_{7},
  {\mathrm{Er}}_{2}{\mathrm{Ti}}_{2}{\mathrm{O}}_{7}$, and
  ${\mathrm{Er}}_{2}{\mathrm{Sn}}_{2}{\mathrm{O}}_{7}$}},}\ }\href {\doibase
  10.1103/PhysRevB.95.094422} {\bibfield  {journal} {\bibinfo  {journal} {Phys.
  Rev. B}\ }\textbf {\bibinfo {volume} {95}},\ \bibinfo {pages} {094422}
  (\bibinfo {year} {2017})}\BibitemShut {NoStop}%
\bibitem [{\citenamefont {Savary}\ \emph {et~al.}(2016)\citenamefont {Savary},
  \citenamefont {Wang}, \citenamefont {Kee}, \citenamefont {Kim}, \citenamefont
  {Yu},\ and\ \citenamefont {Chen}}]{Chen2015}%
  \BibitemOpen
  \bibfield  {author} {\bibinfo {author} {\bibfnamefont {Lucile}\ \bibnamefont
  {Savary}}, \bibinfo {author} {\bibfnamefont {Xiaoqun}\ \bibnamefont {Wang}},
  \bibinfo {author} {\bibfnamefont {Hae-Young}\ \bibnamefont {Kee}}, \bibinfo
  {author} {\bibfnamefont {Yong~Baek}\ \bibnamefont {Kim}}, \bibinfo {author}
  {\bibfnamefont {Yue}\ \bibnamefont {Yu}}, \ and\ \bibinfo {author}
  {\bibfnamefont {Gang}\ \bibnamefont {Chen}},\ }\bibfield  {title} {\enquote
  {\bibinfo {title} {{Quantum spin ice on the breathing pyrochlore lattice}},}\
  }\href {\doibase 10.1103/PhysRevB.94.075146} {\bibfield  {journal} {\bibinfo
  {journal} {Phys. Rev. B}\ }\textbf {\bibinfo {volume} {94}},\ \bibinfo
  {pages} {075146} (\bibinfo {year} {2016})}\BibitemShut {NoStop}%
\bibitem [{\citenamefont {Fennell}\ \emph {et~al.}(2012)\citenamefont
  {Fennell}, \citenamefont {Kenzelmann}, \citenamefont {Roessli}, \citenamefont
  {Haas},\ and\ \citenamefont {Cava}}]{PhysRevLett.109.017201}%
  \BibitemOpen
  \bibfield  {author} {\bibinfo {author} {\bibfnamefont {T.}~\bibnamefont
  {Fennell}}, \bibinfo {author} {\bibfnamefont {M.}~\bibnamefont {Kenzelmann}},
  \bibinfo {author} {\bibfnamefont {B.}~\bibnamefont {Roessli}}, \bibinfo
  {author} {\bibfnamefont {M.~K.}\ \bibnamefont {Haas}}, \ and\ \bibinfo
  {author} {\bibfnamefont {R.~J.}\ \bibnamefont {Cava}},\ }\bibfield  {title}
  {\enquote {\bibinfo {title} {{Power-Law Spin Correlations in the Pyrochlore
  Antiferromagnet ${\mathrm{Tb}}_{2}{\mathrm{Ti}}_{2}{\mathrm{O}}_{7}$}},}\
  }\href {\doibase 10.1103/PhysRevLett.109.017201} {\bibfield  {journal}
  {\bibinfo  {journal} {Phys. Rev. Lett.}\ }\textbf {\bibinfo {volume} {109}},\
  \bibinfo {pages} {017201} (\bibinfo {year} {2012})}\BibitemShut {NoStop}%
\bibitem [{\citenamefont {Yasui}\ \emph {et~al.}(2002)\citenamefont {Yasui},
  \citenamefont {Kanada}, \citenamefont {Ito}, \citenamefont {Harashina},
  \citenamefont {Sato}, \citenamefont {Okumura}, \citenamefont {Kakurai},\ and\
  \citenamefont {Kadowaki}}]{Yasui2002}%
  \BibitemOpen
  \bibfield  {author} {\bibinfo {author} {\bibfnamefont {Yukio}\ \bibnamefont
  {Yasui}}, \bibinfo {author} {\bibfnamefont {Masaki}\ \bibnamefont {Kanada}},
  \bibinfo {author} {\bibfnamefont {Masafumi}\ \bibnamefont {Ito}}, \bibinfo
  {author} {\bibfnamefont {Hiroshi}\ \bibnamefont {Harashina}}, \bibinfo
  {author} {\bibfnamefont {Masatoshi}\ \bibnamefont {Sato}}, \bibinfo {author}
  {\bibfnamefont {Hajime}\ \bibnamefont {Okumura}}, \bibinfo {author}
  {\bibfnamefont {Kazuhisa}\ \bibnamefont {Kakurai}}, \ and\ \bibinfo {author}
  {\bibfnamefont {Hiroaki}\ \bibnamefont {Kadowaki}},\ }\bibfield  {title}
  {\enquote {\bibinfo {title} {{Static Correlation and Dynamical Properties of
  Tb$^{3+}$-moments in Tb$_2$Ti$_2$O$_7$ –Neutron Scattering Study–}},}\
  }\href {\doibase 10.1143/JPSJ.71.599} {\bibfield  {journal} {\bibinfo
  {journal} {Journal of the Physical Society of Japan}\ }\textbf {\bibinfo
  {volume} {71}},\ \bibinfo {pages} {599--606} (\bibinfo {year}
  {2002})}\BibitemShut {NoStop}%
\bibitem [{\citenamefont {Gardner}\ \emph {et~al.}(2001)\citenamefont
  {Gardner}, \citenamefont {Gaulin}, \citenamefont {Berlinsky}, \citenamefont
  {Waldron}, \citenamefont {Dunsiger}, \citenamefont {Raju},\ and\
  \citenamefont {Greedan}}]{PhysRevB.64.224416}%
  \BibitemOpen
  \bibfield  {author} {\bibinfo {author} {\bibfnamefont {J.~S.}\ \bibnamefont
  {Gardner}}, \bibinfo {author} {\bibfnamefont {B.~D.}\ \bibnamefont {Gaulin}},
  \bibinfo {author} {\bibfnamefont {A.~J.}\ \bibnamefont {Berlinsky}}, \bibinfo
  {author} {\bibfnamefont {P.}~\bibnamefont {Waldron}}, \bibinfo {author}
  {\bibfnamefont {S.~R.}\ \bibnamefont {Dunsiger}}, \bibinfo {author}
  {\bibfnamefont {N.~P.}\ \bibnamefont {Raju}}, \ and\ \bibinfo {author}
  {\bibfnamefont {J.~E.}\ \bibnamefont {Greedan}},\ }\bibfield  {title}
  {\enquote {\bibinfo {title} {{Neutron scattering studies of the cooperative
  paramagnet pyrochlore
  ${\mathrm{Tb}}_{2}{\mathrm{Ti}}_{2}{\mathrm{O}}_{7}$}},}\ }\href {\doibase
  10.1103/PhysRevB.64.224416} {\bibfield  {journal} {\bibinfo  {journal} {Phys.
  Rev. B}\ }\textbf {\bibinfo {volume} {64}},\ \bibinfo {pages} {224416}
  (\bibinfo {year} {2001})}\BibitemShut {NoStop}%
\bibitem [{\citenamefont {Hao}\ \emph {et~al.}(2014)\citenamefont {Hao},
  \citenamefont {Day},\ and\ \citenamefont {Gingras}}]{PhysRevB.90.214430}%
  \BibitemOpen
  \bibfield  {author} {\bibinfo {author} {\bibfnamefont {Zhihao}\ \bibnamefont
  {Hao}}, \bibinfo {author} {\bibfnamefont {Alexandre G.~R.}\ \bibnamefont
  {Day}}, \ and\ \bibinfo {author} {\bibfnamefont {Michel J.~P.}\ \bibnamefont
  {Gingras}},\ }\bibfield  {title} {\enquote {\bibinfo {title} {{Bosonic
  many-body theory of quantum spin ice}},}\ }\href {\doibase
  10.1103/PhysRevB.90.214430} {\bibfield  {journal} {\bibinfo  {journal} {Phys.
  Rev. B}\ }\textbf {\bibinfo {volume} {90}},\ \bibinfo {pages} {214430}
  (\bibinfo {year} {2014})}\BibitemShut {NoStop}%
\bibitem [{\citenamefont {Chang}\ \emph {et~al.}(2012)\citenamefont {Chang},
  \citenamefont {Onoda}, \citenamefont {Su}, \citenamefont {Kao}, \citenamefont
  {Tsuei}, \citenamefont {Yasui}, \citenamefont {Kakurai},\ and\ \citenamefont
  {Lees}}]{Chang2012}%
  \BibitemOpen
  \bibfield  {author} {\bibinfo {author} {\bibfnamefont {Lieh-Jeng}\
  \bibnamefont {Chang}}, \bibinfo {author} {\bibfnamefont {Shigeki}\
  \bibnamefont {Onoda}}, \bibinfo {author} {\bibfnamefont {Yixi}\ \bibnamefont
  {Su}}, \bibinfo {author} {\bibfnamefont {Ying-Jer}\ \bibnamefont {Kao}},
  \bibinfo {author} {\bibfnamefont {Ku-Ding}\ \bibnamefont {Tsuei}}, \bibinfo
  {author} {\bibfnamefont {Yukio}\ \bibnamefont {Yasui}}, \bibinfo {author}
  {\bibfnamefont {Kazuhisa}\ \bibnamefont {Kakurai}}, \ and\ \bibinfo {author}
  {\bibfnamefont {Martin~Richard}\ \bibnamefont {Lees}},\ }\bibfield  {title}
  {\enquote {\bibinfo {title} {{Higgs transition from a magnetic Coulomb liquid
  to a ferromagnet in Yb$_2$Ti$_2$O$_7$}},}\ }\href {\doibase
  10.1038/ncomms1989} {\bibfield  {journal} {\bibinfo  {journal} {Nature
  Communications}\ }\textbf {\bibinfo {volume} {3}},\ \bibinfo {pages} {992}
  (\bibinfo {year} {2012})}\BibitemShut {NoStop}%
\bibitem [{\citenamefont {Kimura}\ \emph {et~al.}(2013)\citenamefont {Kimura},
  \citenamefont {Nakatsuji}, \citenamefont {Wen}, \citenamefont {Broholm},
  \citenamefont {Stone}, \citenamefont {Nishibori},\ and\ \citenamefont
  {Sawa}}]{Kimura2012}%
  \BibitemOpen
  \bibfield  {author} {\bibinfo {author} {\bibfnamefont {K.}~\bibnamefont
  {Kimura}}, \bibinfo {author} {\bibfnamefont {K.}~\bibnamefont {Nakatsuji}},
  \bibinfo {author} {\bibfnamefont {J-J.}\ \bibnamefont {Wen}}, \bibinfo
  {author} {\bibfnamefont {C.}~\bibnamefont {Broholm}}, \bibinfo {author}
  {\bibfnamefont {M.B.}\ \bibnamefont {Stone}}, \bibinfo {author}
  {\bibfnamefont {E.}~\bibnamefont {Nishibori}}, \ and\ \bibinfo {author}
  {\bibfnamefont {H.}~\bibnamefont {Sawa}},\ }\bibfield  {title} {\enquote
  {\bibinfo {title} {{Quantum fluctuations in spin-ice-like
  Pr$_2$Zr$_2$O$_7$}},}\ }\href {\doibase 10.1038/ncomms2914} {\bibfield
  {journal} {\bibinfo  {journal} {Nature Communications}\ }\textbf {\bibinfo
  {volume} {4}},\ \bibinfo {pages} {2914} (\bibinfo {year} {2013})}\BibitemShut
  {NoStop}%
\bibitem [{\citenamefont {Gardner}\ \emph {et~al.}(2010)\citenamefont
  {Gardner}, \citenamefont {Gingras},\ and\ \citenamefont
  {Greedan}}]{RevModPhys.82.53}%
  \BibitemOpen
  \bibfield  {author} {\bibinfo {author} {\bibfnamefont {Jason~S.}\
  \bibnamefont {Gardner}}, \bibinfo {author} {\bibfnamefont {Michel J.~P.}\
  \bibnamefont {Gingras}}, \ and\ \bibinfo {author} {\bibfnamefont {John~E.}\
  \bibnamefont {Greedan}},\ }\bibfield  {title} {\enquote {\bibinfo {title}
  {{Magnetic pyrochlore oxides}},}\ }\href {\doibase 10.1103/RevModPhys.82.53}
  {\bibfield  {journal} {\bibinfo  {journal} {Rev. Mod. Phys.}\ }\textbf
  {\bibinfo {volume} {82}},\ \bibinfo {pages} {53--107} (\bibinfo {year}
  {2010})}\BibitemShut {NoStop}%
\bibitem [{\citenamefont {Lhotel}\ \emph {et~al.}(2014)\citenamefont {Lhotel},
  \citenamefont {Giblin}, \citenamefont {Lees}, \citenamefont {Balakrishnan},
  \citenamefont {Chang},\ and\ \citenamefont {Yasui}}]{Lhotel2014}%
  \BibitemOpen
  \bibfield  {author} {\bibinfo {author} {\bibfnamefont {E.}~\bibnamefont
  {Lhotel}}, \bibinfo {author} {\bibfnamefont {S.~R.}\ \bibnamefont {Giblin}},
  \bibinfo {author} {\bibfnamefont {M.~R.}\ \bibnamefont {Lees}}, \bibinfo
  {author} {\bibfnamefont {G.}~\bibnamefont {Balakrishnan}}, \bibinfo {author}
  {\bibfnamefont {L.~J.}\ \bibnamefont {Chang}}, \ and\ \bibinfo {author}
  {\bibfnamefont {Y.}~\bibnamefont {Yasui}},\ }\bibfield  {title} {\enquote
  {\bibinfo {title} {{First-order magnetic transition in
  ${\mathrm{Yb}}_{2}{\mathrm{Ti}}_{2}{\mathrm{O}}_{7}$}},}\ }\href {\doibase
  10.1103/PhysRevB.89.224419} {\bibfield  {journal} {\bibinfo  {journal} {Phys.
  Rev. B}\ }\textbf {\bibinfo {volume} {89}},\ \bibinfo {pages} {224419}
  (\bibinfo {year} {2014})}\BibitemShut {NoStop}%
\bibitem [{\citenamefont {Chang}\ \emph {et~al.}(2014)\citenamefont {Chang},
  \citenamefont {Lees}, \citenamefont {Watanabe}, \citenamefont {Hillier},
  \citenamefont {Yasui},\ and\ \citenamefont {Onoda}}]{Chang2014}%
  \BibitemOpen
  \bibfield  {author} {\bibinfo {author} {\bibfnamefont {Lieh-Jeng}\
  \bibnamefont {Chang}}, \bibinfo {author} {\bibfnamefont {Martin~R.}\
  \bibnamefont {Lees}}, \bibinfo {author} {\bibfnamefont {Isao}\ \bibnamefont
  {Watanabe}}, \bibinfo {author} {\bibfnamefont {Adrian~D.}\ \bibnamefont
  {Hillier}}, \bibinfo {author} {\bibfnamefont {Yukio}\ \bibnamefont {Yasui}},
  \ and\ \bibinfo {author} {\bibfnamefont {Shigeki}\ \bibnamefont {Onoda}},\
  }\bibfield  {title} {\enquote {\bibinfo {title} {{Static magnetic moments
  revealed by muon spin relaxation and thermodynamic measurements in the
  quantum spin ice ${\text{Yb}}_{2}{\text{Ti}}_{2}{\text{O}}_{7}$}},}\ }\href
  {\doibase 10.1103/PhysRevB.89.184416} {\bibfield  {journal} {\bibinfo
  {journal} {Phys. Rev. B}\ }\textbf {\bibinfo {volume} {89}},\ \bibinfo
  {pages} {184416} (\bibinfo {year} {2014})}\BibitemShut {NoStop}%
\bibitem [{\citenamefont {Yasui}\ \emph {et~al.}(2003)\citenamefont {Yasui},
  \citenamefont {Soda}, \citenamefont {Iikubo}, \citenamefont {Ito},
  \citenamefont {Sato}, \citenamefont {Hamaguchi}, \citenamefont {Matsushita},
  \citenamefont {Wada}, \citenamefont {Takeuchi}, \citenamefont {Aso},\ and\
  \citenamefont {Kakurai}}]{Yasui2003}%
  \BibitemOpen
  \bibfield  {author} {\bibinfo {author} {\bibfnamefont {Yukio}\ \bibnamefont
  {Yasui}}, \bibinfo {author} {\bibfnamefont {Minoru}\ \bibnamefont {Soda}},
  \bibinfo {author} {\bibfnamefont {Satoshi}\ \bibnamefont {Iikubo}}, \bibinfo
  {author} {\bibfnamefont {Masafumi}\ \bibnamefont {Ito}}, \bibinfo {author}
  {\bibfnamefont {Masatoshi}\ \bibnamefont {Sato}}, \bibinfo {author}
  {\bibfnamefont {Nobuko}\ \bibnamefont {Hamaguchi}}, \bibinfo {author}
  {\bibfnamefont {Taku}\ \bibnamefont {Matsushita}}, \bibinfo {author}
  {\bibfnamefont {Nobuo}\ \bibnamefont {Wada}}, \bibinfo {author}
  {\bibfnamefont {Tetsuya}\ \bibnamefont {Takeuchi}}, \bibinfo {author}
  {\bibfnamefont {Naofumi}\ \bibnamefont {Aso}}, \ and\ \bibinfo {author}
  {\bibfnamefont {Kazuhisa}\ \bibnamefont {Kakurai}},\ }\bibfield  {title}
  {\enquote {\bibinfo {title} {{Ferromagnetic Transition of Pyrochlore Compound
  Yb$_2$Ti$_2$O$_7$}},}\ }\href {\doibase 10.1143/JPSJ.72.3014} {\bibfield
  {journal} {\bibinfo  {journal} {Journal of the Physical Society of Japan}\
  }\textbf {\bibinfo {volume} {72}},\ \bibinfo {pages} {3014--3015} (\bibinfo
  {year} {2003})}\BibitemShut {NoStop}%
\bibitem [{\citenamefont {Ross}\ \emph {et~al.}(2011)\citenamefont {Ross},
  \citenamefont {Savary}, \citenamefont {Gaulin},\ and\ \citenamefont
  {Balents}}]{Ross11}%
  \BibitemOpen
  \bibfield  {author} {\bibinfo {author} {\bibfnamefont {Kate}\ \bibnamefont
  {Ross}}, \bibinfo {author} {\bibfnamefont {Lucile}\ \bibnamefont {Savary}},
  \bibinfo {author} {\bibfnamefont {Bruce}\ \bibnamefont {Gaulin}}, \ and\
  \bibinfo {author} {\bibfnamefont {Leon}\ \bibnamefont {Balents}},\ }\bibfield
   {title} {\enquote {\bibinfo {title} {{Quantum Excitations in Quantum Spin
  Ice}},}\ }\href {\doibase 10.1103/PhysRevX.1.021002} {\bibfield  {journal}
  {\bibinfo  {journal} {Phys. Rev. X}\ }\textbf {\bibinfo {volume} {1}},\
  \bibinfo {pages} {021002} (\bibinfo {year} {2011})}\BibitemShut {NoStop}%
\bibitem [{\citenamefont {Shannon}\ \emph {et~al.}(2012)\citenamefont
  {Shannon}, \citenamefont {Sikora}, \citenamefont {Pollmann}, \citenamefont
  {Penc},\ and\ \citenamefont {Fulde}}]{Shannon12}%
  \BibitemOpen
  \bibfield  {author} {\bibinfo {author} {\bibfnamefont {Nic}\ \bibnamefont
  {Shannon}}, \bibinfo {author} {\bibfnamefont {Olga}\ \bibnamefont {Sikora}},
  \bibinfo {author} {\bibfnamefont {Frank}\ \bibnamefont {Pollmann}}, \bibinfo
  {author} {\bibfnamefont {Karlo}\ \bibnamefont {Penc}}, \ and\ \bibinfo
  {author} {\bibfnamefont {Peter}\ \bibnamefont {Fulde}},\ }\bibfield  {title}
  {\enquote {\bibinfo {title} {{Quantum Ice: A Quantum Monte Carlo Study}},}\
  }\href {\doibase 10.1103/PhysRevLett.108.067204} {\bibfield  {journal}
  {\bibinfo  {journal} {Phys. Rev. Lett.}\ }\textbf {\bibinfo {volume} {108}},\
  \bibinfo {pages} {067204} (\bibinfo {year} {2012})}\BibitemShut {NoStop}%
\bibitem [{\citenamefont {Goswami}\ \emph {et~al.}(2016)\citenamefont
  {Goswami}, \citenamefont {Roy},\ and\ \citenamefont
  {Das~Sarma}}]{Goswami2016}%
  \BibitemOpen
  \bibfield  {author} {\bibinfo {author} {\bibfnamefont {Pallab}\ \bibnamefont
  {Goswami}}, \bibinfo {author} {\bibfnamefont {Bitan}\ \bibnamefont {Roy}}, \
  and\ \bibinfo {author} {\bibfnamefont {Sankar}\ \bibnamefont {Das~Sarma}},\
  }\bibfield  {title} {\enquote {\bibinfo {title} {{Itinerant spin ice order,
  Weyl metal, and anomalous Hall effect in Pr$_2$Ir$_2$O$_7$}},}\ }\href@noop
  {} {\bibfield  {journal} {\bibinfo  {journal} {ArXiv:1603.02273}\ } (\bibinfo
  {year} {2016})}\BibitemShut {NoStop}%
\bibitem [{\citenamefont {Arpino}\ \emph {et~al.}(2017)\citenamefont {Arpino},
  \citenamefont {Trump}, \citenamefont {Scheie}, \citenamefont {McQueen},\ and\
  \citenamefont {Koohpayeh}}]{PhysRevB.95.094407}%
  \BibitemOpen
  \bibfield  {author} {\bibinfo {author} {\bibfnamefont {K.~E.}\ \bibnamefont
  {Arpino}}, \bibinfo {author} {\bibfnamefont {B.~A.}\ \bibnamefont {Trump}},
  \bibinfo {author} {\bibfnamefont {A.~O.}\ \bibnamefont {Scheie}}, \bibinfo
  {author} {\bibfnamefont {T.~M.}\ \bibnamefont {McQueen}}, \ and\ \bibinfo
  {author} {\bibfnamefont {S.~M.}\ \bibnamefont {Koohpayeh}},\ }\bibfield
  {title} {\enquote {\bibinfo {title} {{Impact of stoichiometry of
  ${\mathrm{Yb}}_{2}{\mathrm{Ti}}_{2}{\mathrm{O}}_{7}$ on its physical
  properties}},}\ }\href {\doibase 10.1103/PhysRevB.95.094407} {\bibfield
  {journal} {\bibinfo  {journal} {Phys. Rev. B}\ }\textbf {\bibinfo {volume}
  {95}},\ \bibinfo {pages} {094407} (\bibinfo {year} {2017})}\BibitemShut
  {NoStop}%
\bibitem [{\citenamefont {Chen}(2016)}]{PhysRevB.94.205107}%
  \BibitemOpen
  \bibfield  {author} {\bibinfo {author} {\bibfnamefont {Gang}\ \bibnamefont
  {Chen}},\ }\bibfield  {title} {\enquote {\bibinfo {title} {{``Magnetic
  monopole'' condensation of the pyrochlore ice U(1) quantum spin liquid:
  Application to ${\mathrm{Pr}}_{2}{\mathrm{Ir}}_{2}{\mathrm{O}}_{7}$ and
  ${\mathrm{Yb}}_{2}{\mathrm{Ti}}_{2}{\mathrm{O}}_{7}$}},}\ }\href {\doibase
  10.1103/PhysRevB.94.205107} {\bibfield  {journal} {\bibinfo  {journal} {Phys.
  Rev. B}\ }\textbf {\bibinfo {volume} {94}},\ \bibinfo {pages} {205107}
  (\bibinfo {year} {2016})}\BibitemShut {NoStop}%
\bibitem [{\citenamefont {Wen}\ \emph {et~al.}(2017)\citenamefont {Wen},
  \citenamefont {Koohpayeh}, \citenamefont {Ross}, \citenamefont {Trump},
  \citenamefont {McQueen}, \citenamefont {Kimura}, \citenamefont {Nakatsuji},
  \citenamefont {Qiu}, \citenamefont {Pajerowski}, \citenamefont {Copley},\
  and\ \citenamefont {Broholm}}]{PhysRevLett.118.107206}%
  \BibitemOpen
  \bibfield  {author} {\bibinfo {author} {\bibfnamefont {J.-J.}\ \bibnamefont
  {Wen}}, \bibinfo {author} {\bibfnamefont {S.~M.}\ \bibnamefont {Koohpayeh}},
  \bibinfo {author} {\bibfnamefont {K.~A.}\ \bibnamefont {Ross}}, \bibinfo
  {author} {\bibfnamefont {B.~A.}\ \bibnamefont {Trump}}, \bibinfo {author}
  {\bibfnamefont {T.~M.}\ \bibnamefont {McQueen}}, \bibinfo {author}
  {\bibfnamefont {K.}~\bibnamefont {Kimura}}, \bibinfo {author} {\bibfnamefont
  {S.}~\bibnamefont {Nakatsuji}}, \bibinfo {author} {\bibfnamefont
  {Y.}~\bibnamefont {Qiu}}, \bibinfo {author} {\bibfnamefont {D.~M.}\
  \bibnamefont {Pajerowski}}, \bibinfo {author} {\bibfnamefont {J.~R.~D.}\
  \bibnamefont {Copley}}, \ and\ \bibinfo {author} {\bibfnamefont {C.~L.}\
  \bibnamefont {Broholm}},\ }\bibfield  {title} {\enquote {\bibinfo {title}
  {{Disordered Route to the Coulomb Quantum Spin Liquid: Random Transverse
  Fields on Spin Ice in
  ${\mathrm{Pr}}_{2}{\mathrm{Zr}}_{2}{\mathrm{O}}_{7}$}},}\ }\href {\doibase
  10.1103/PhysRevLett.118.107206} {\bibfield  {journal} {\bibinfo  {journal}
  {Phys. Rev. Lett.}\ }\textbf {\bibinfo {volume} {118}},\ \bibinfo {pages}
  {107206} (\bibinfo {year} {2017})}\BibitemShut {NoStop}%
\bibitem [{\citenamefont {Chen}(2017)}]{Gangchen201706}%
  \BibitemOpen
  \bibfield  {author} {\bibinfo {author} {\bibfnamefont {Gang}\ \bibnamefont
  {Chen}},\ }\bibfield  {title} {\enquote {\bibinfo {title} {{What does
  inelastic neutron scattering measure in quantum spin ices?}}}\ }\href@noop {}
  {\bibfield  {journal} {\bibinfo  {journal} {ArXiv:1706.04333}\ } (\bibinfo
  {year} {2017})}\BibitemShut {NoStop}%
\bibitem [{\citenamefont {MacLaughlin}\ \emph {et~al.}(2015)\citenamefont
  {MacLaughlin}, \citenamefont {Bernal}, \citenamefont {Shu}, \citenamefont
  {Ishikawa}, \citenamefont {Matsumoto}, \citenamefont {Wen}, \citenamefont
  {Mourigal}, \citenamefont {Stock}, \citenamefont {Ehlers}, \citenamefont
  {Broholm}, \citenamefont {Machida}, \citenamefont {Kimura}, \citenamefont
  {Nakatsuji}, \citenamefont {Shimura},\ and\ \citenamefont
  {Sakakibara}}]{PhysRevB.92.054432}%
  \BibitemOpen
  \bibfield  {author} {\bibinfo {author} {\bibfnamefont {D.~E.}\ \bibnamefont
  {MacLaughlin}}, \bibinfo {author} {\bibfnamefont {O.~O.}\ \bibnamefont
  {Bernal}}, \bibinfo {author} {\bibfnamefont {Lei}\ \bibnamefont {Shu}},
  \bibinfo {author} {\bibfnamefont {Jun}\ \bibnamefont {Ishikawa}}, \bibinfo
  {author} {\bibfnamefont {Yosuke}\ \bibnamefont {Matsumoto}}, \bibinfo
  {author} {\bibfnamefont {J.-J.}\ \bibnamefont {Wen}}, \bibinfo {author}
  {\bibfnamefont {M.}~\bibnamefont {Mourigal}}, \bibinfo {author}
  {\bibfnamefont {C.}~\bibnamefont {Stock}}, \bibinfo {author} {\bibfnamefont
  {G.}~\bibnamefont {Ehlers}}, \bibinfo {author} {\bibfnamefont {C.~L.}\
  \bibnamefont {Broholm}}, \bibinfo {author} {\bibfnamefont {Yo}~\bibnamefont
  {Machida}}, \bibinfo {author} {\bibfnamefont {Kenta}\ \bibnamefont {Kimura}},
  \bibinfo {author} {\bibfnamefont {Satoru}\ \bibnamefont {Nakatsuji}},
  \bibinfo {author} {\bibfnamefont {Yasuyuki}\ \bibnamefont {Shimura}}, \ and\
  \bibinfo {author} {\bibfnamefont {Toshiro}\ \bibnamefont {Sakakibara}},\
  }\bibfield  {title} {\enquote {\bibinfo {title} {{Unstable spin-ice order in
  the stuffed metallic pyrochlore
  ${\mathrm{Pr}}_{2+x}{\mathrm{Ir}}_{2\ensuremath{-}x}{\mathrm{O}}_{7\ensuremath{-}\ensuremath{\delta}}$}},}\
  }\href {\doibase 10.1103/PhysRevB.92.054432} {\bibfield  {journal} {\bibinfo
  {journal} {Phys. Rev. B}\ }\textbf {\bibinfo {volume} {92}},\ \bibinfo
  {pages} {054432} (\bibinfo {year} {2015})}\BibitemShut {NoStop}%
\bibitem [{\citenamefont {Fu}\ \emph {et~al.}(2017)\citenamefont {Fu},
  \citenamefont {Rau}, \citenamefont {Gingras},\ and\ \citenamefont
  {Perkins}}]{fu2017fingerprints}%
  \BibitemOpen
  \bibfield  {author} {\bibinfo {author} {\bibfnamefont {Jianlong}\
  \bibnamefont {Fu}}, \bibinfo {author} {\bibfnamefont {Jeffrey~G}\
  \bibnamefont {Rau}}, \bibinfo {author} {\bibfnamefont {Michel~JP}\
  \bibnamefont {Gingras}}, \ and\ \bibinfo {author} {\bibfnamefont {Natalia~B}\
  \bibnamefont {Perkins}},\ }\bibfield  {title} {\enquote {\bibinfo {title}
  {Fingerprints of quantum spin ice in raman scattering},}\ }\href@noop {}
  {\bibfield  {journal} {\bibinfo  {journal} {arXiv preprint arXiv:1703.03836}\
  } (\bibinfo {year} {2017})}\BibitemShut {NoStop}%
\bibitem [{\citenamefont {Benton}\ \emph {et~al.}(2012)\citenamefont {Benton},
  \citenamefont {Sikora},\ and\ \citenamefont {Shannon}}]{PhysRevB.86.075154}%
  \BibitemOpen
  \bibfield  {author} {\bibinfo {author} {\bibfnamefont {Owen}\ \bibnamefont
  {Benton}}, \bibinfo {author} {\bibfnamefont {Olga}\ \bibnamefont {Sikora}}, \
  and\ \bibinfo {author} {\bibfnamefont {Nic}\ \bibnamefont {Shannon}},\
  }\bibfield  {title} {\enquote {\bibinfo {title} {{Seeing the light:
  Experimental signatures of emergent electromagnetism in a quantum spin
  ice}},}\ }\href {\doibase 10.1103/PhysRevB.86.075154} {\bibfield  {journal}
  {\bibinfo  {journal} {Phys. Rev. B}\ }\textbf {\bibinfo {volume} {86}},\
  \bibinfo {pages} {075154} (\bibinfo {year} {2012})}\BibitemShut {NoStop}%
\bibitem [{\citenamefont {Jaubert}\ \emph {et~al.}(2015)\citenamefont
  {Jaubert}, \citenamefont {Benton}, \citenamefont {Rau}, \citenamefont
  {Oitmaa}, \citenamefont {Singh}, \citenamefont {Shannon},\ and\ \citenamefont
  {Gingras}}]{PhysRevLett.115.267208}%
  \BibitemOpen
  \bibfield  {author} {\bibinfo {author} {\bibfnamefont {L.~D.~C.}\
  \bibnamefont {Jaubert}}, \bibinfo {author} {\bibfnamefont {Owen}\
  \bibnamefont {Benton}}, \bibinfo {author} {\bibfnamefont {Jeffrey~G.}\
  \bibnamefont {Rau}}, \bibinfo {author} {\bibfnamefont {J.}~\bibnamefont
  {Oitmaa}}, \bibinfo {author} {\bibfnamefont {R.~R.~P.}\ \bibnamefont
  {Singh}}, \bibinfo {author} {\bibfnamefont {Nic}\ \bibnamefont {Shannon}}, \
  and\ \bibinfo {author} {\bibfnamefont {Michel J.~P.}\ \bibnamefont
  {Gingras}},\ }\bibfield  {title} {\enquote {\bibinfo {title} {{Are Multiphase
  Competition and Order by Disorder the Keys to Understanding
  ${\mathrm{Yb}}_{2}{\mathrm{Ti}}_{2}{\mathrm{O}}_{7}$?}}}\ }\href {\doibase
  10.1103/PhysRevLett.115.267208} {\bibfield  {journal} {\bibinfo  {journal}
  {Phys. Rev. Lett.}\ }\textbf {\bibinfo {volume} {115}},\ \bibinfo {pages}
  {267208} (\bibinfo {year} {2015})}\BibitemShut {NoStop}%
\bibitem [{\citenamefont {Applegate}\ \emph {et~al.}(2012)\citenamefont
  {Applegate}, \citenamefont {Hayre}, \citenamefont {Singh}, \citenamefont
  {Lin}, \citenamefont {Day},\ and\ \citenamefont
  {Gingras}}]{PhysRevLett.109.097205}%
  \BibitemOpen
  \bibfield  {author} {\bibinfo {author} {\bibfnamefont {R.}~\bibnamefont
  {Applegate}}, \bibinfo {author} {\bibfnamefont {N.~R.}\ \bibnamefont
  {Hayre}}, \bibinfo {author} {\bibfnamefont {R.~R.~P.}\ \bibnamefont {Singh}},
  \bibinfo {author} {\bibfnamefont {T.}~\bibnamefont {Lin}}, \bibinfo {author}
  {\bibfnamefont {A.~G.~R.}\ \bibnamefont {Day}}, \ and\ \bibinfo {author}
  {\bibfnamefont {M.~J.~P.}\ \bibnamefont {Gingras}},\ }\bibfield  {title}
  {\enquote {\bibinfo {title} {{Vindication of
  ${\mathrm{Yb}}_{2}{\mathrm{Ti}}_{2}{\mathrm{O}}_{7}$ as a Model Exchange
  Quantum Spin Ice}},}\ }\href {\doibase 10.1103/PhysRevLett.109.097205}
  {\bibfield  {journal} {\bibinfo  {journal} {Phys. Rev. Lett.}\ }\textbf
  {\bibinfo {volume} {109}},\ \bibinfo {pages} {097205} (\bibinfo {year}
  {2012})}\BibitemShut {NoStop}%
\bibitem [{\citenamefont {Taillefumier}\ \emph {et~al.}(2017)\citenamefont
  {Taillefumier}, \citenamefont {Benton}, \citenamefont {Yan}, \citenamefont
  {Jaubert},\ and\ \citenamefont {Shannon}}]{Frustrating1705}%
  \BibitemOpen
  \bibfield  {author} {\bibinfo {author} {\bibfnamefont {Mathieu}\ \bibnamefont
  {Taillefumier}}, \bibinfo {author} {\bibfnamefont {Owen}\ \bibnamefont
  {Benton}}, \bibinfo {author} {\bibfnamefont {Han}\ \bibnamefont {Yan}},
  \bibinfo {author} {\bibfnamefont {Ludovic}\ \bibnamefont {Jaubert}}, \ and\
  \bibinfo {author} {\bibfnamefont {Nic}\ \bibnamefont {Shannon}},\ }\bibfield
  {title} {\enquote {\bibinfo {title} {{Frustrating quantum spin ice: a tale of
  three spin liquids, and hidden order}},}\ }\href@noop {} {\bibfield
  {journal} {\bibinfo  {journal} {ArXiv:1705.00148}\ } (\bibinfo {year}
  {2017})}\BibitemShut {NoStop}%
\bibitem [{\citenamefont {Dunsiger}\ \emph {et~al.}(2011)\citenamefont
  {Dunsiger}, \citenamefont {Aczel}, \citenamefont {Arguello}, \citenamefont
  {Dabkowska}, \citenamefont {Dabkowski}, \citenamefont {Du}, \citenamefont
  {Goko}, \citenamefont {Javanparast}, \citenamefont {Lin}, \citenamefont
  {Ning}, \citenamefont {Noad}, \citenamefont {Singh}, \citenamefont
  {Williams}, \citenamefont {Uemura}, \citenamefont {Gingras},\ and\
  \citenamefont {Luke}}]{PhysRevLett.107.207207}%
  \BibitemOpen
  \bibfield  {author} {\bibinfo {author} {\bibfnamefont {S.~R.}\ \bibnamefont
  {Dunsiger}}, \bibinfo {author} {\bibfnamefont {A.~A.}\ \bibnamefont {Aczel}},
  \bibinfo {author} {\bibfnamefont {C.}~\bibnamefont {Arguello}}, \bibinfo
  {author} {\bibfnamefont {H.}~\bibnamefont {Dabkowska}}, \bibinfo {author}
  {\bibfnamefont {A.}~\bibnamefont {Dabkowski}}, \bibinfo {author}
  {\bibfnamefont {M.-H.}\ \bibnamefont {Du}}, \bibinfo {author} {\bibfnamefont
  {T.}~\bibnamefont {Goko}}, \bibinfo {author} {\bibfnamefont {B.}~\bibnamefont
  {Javanparast}}, \bibinfo {author} {\bibfnamefont {T.}~\bibnamefont {Lin}},
  \bibinfo {author} {\bibfnamefont {F.~L.}\ \bibnamefont {Ning}}, \bibinfo
  {author} {\bibfnamefont {H.~M.~L.}\ \bibnamefont {Noad}}, \bibinfo {author}
  {\bibfnamefont {D.~J.}\ \bibnamefont {Singh}}, \bibinfo {author}
  {\bibfnamefont {T.~J.}\ \bibnamefont {Williams}}, \bibinfo {author}
  {\bibfnamefont {Y.~J.}\ \bibnamefont {Uemura}}, \bibinfo {author}
  {\bibfnamefont {M.~J.~P.}\ \bibnamefont {Gingras}}, \ and\ \bibinfo {author}
  {\bibfnamefont {G.~M.}\ \bibnamefont {Luke}},\ }\bibfield  {title} {\enquote
  {\bibinfo {title} {Spin ice: Magnetic excitations without monopole signatures
  using muon spin rotation},}\ }\href {\doibase 10.1103/PhysRevLett.107.207207}
  {\bibfield  {journal} {\bibinfo  {journal} {Phys. Rev. Lett.}\ }\textbf
  {\bibinfo {volume} {107}},\ \bibinfo {pages} {207207} (\bibinfo {year}
  {2011})}\BibitemShut {NoStop}%
\bibitem [{\citenamefont {Sibille}\ \emph {et~al.}(2015)\citenamefont
  {Sibille}, \citenamefont {Lhotel}, \citenamefont {Pomjakushin}, \citenamefont
  {Baines}, \citenamefont {Fennell},\ and\ \citenamefont
  {Kenzelmann}}]{PhysRevLett.115.097202}%
  \BibitemOpen
  \bibfield  {author} {\bibinfo {author} {\bibfnamefont {Romain}\ \bibnamefont
  {Sibille}}, \bibinfo {author} {\bibfnamefont {Elsa}\ \bibnamefont {Lhotel}},
  \bibinfo {author} {\bibfnamefont {Vladimir}\ \bibnamefont {Pomjakushin}},
  \bibinfo {author} {\bibfnamefont {Chris}\ \bibnamefont {Baines}}, \bibinfo
  {author} {\bibfnamefont {Tom}\ \bibnamefont {Fennell}}, \ and\ \bibinfo
  {author} {\bibfnamefont {Michel}\ \bibnamefont {Kenzelmann}},\ }\bibfield
  {title} {\enquote {\bibinfo {title} {{Candidate Quantum Spin Liquid in the
  ${\mathrm{Ce}}^{3+}$ Pyrochlore Stannate
  ${\mathrm{Ce}}_{2}{\mathrm{Sn}}_{2}{\mathrm{O}}_{7}$}},}\ }\href {\doibase
  10.1103/PhysRevLett.115.097202} {\bibfield  {journal} {\bibinfo  {journal}
  {Phys. Rev. Lett.}\ }\textbf {\bibinfo {volume} {115}},\ \bibinfo {pages}
  {097202} (\bibinfo {year} {2015})}\BibitemShut {NoStop}%
\bibitem [{MIS()}]{MISC}%
  \BibitemOpen
  \href@noop {} {}\bibinfo {note} {\textnormal{For the XXZ model, AFM$_0$ is a
  ferromagnetic state. But in the local coordinate system for pyrochlore spin
  ice materials, it is antiferromagnetic. So we adopt the notation of AFM
  here.}}\BibitemShut {Stop}%
\bibitem [{\citenamefont {de~Picciotto}\ \emph {et~al.}(1997)\citenamefont
  {de~Picciotto}, \citenamefont {Reznikov}, \citenamefont {Heiblum},
  \citenamefont {Umansky}, \citenamefont {Bunin},\ and\ \citenamefont
  {Mahalu}}]{nature_fqh}%
  \BibitemOpen
  \bibfield  {author} {\bibinfo {author} {\bibfnamefont {R.}~\bibnamefont
  {de~Picciotto}}, \bibinfo {author} {\bibfnamefont {M.}~\bibnamefont
  {Reznikov}}, \bibinfo {author} {\bibfnamefont {M.}~\bibnamefont {Heiblum}},
  \bibinfo {author} {\bibfnamefont {V.}~\bibnamefont {Umansky}}, \bibinfo
  {author} {\bibfnamefont {G.}~\bibnamefont {Bunin}}, \ and\ \bibinfo {author}
  {\bibfnamefont {D.}~\bibnamefont {Mahalu}},\ }\bibfield  {title} {\enquote
  {\bibinfo {title} {Direct observation of a fractional charge},}\ }\href
  {\doibase 10.1038/38241} {\bibfield  {journal} {\bibinfo  {journal} {Nature}\
  }\textbf {\bibinfo {volume} {389}},\ \bibinfo {pages} {162--164} (\bibinfo
  {year} {1997})}\BibitemShut {NoStop}%
\bibitem [{\citenamefont {Goldman}\ and\ \citenamefont
  {Su}(1995)}]{Goldman1010}%
  \BibitemOpen
  \bibfield  {author} {\bibinfo {author} {\bibfnamefont {V.~J.}\ \bibnamefont
  {Goldman}}\ and\ \bibinfo {author} {\bibfnamefont {B.}~\bibnamefont {Su}},\
  }\bibfield  {title} {\enquote {\bibinfo {title} {Resonant tunneling in the
  quantum hall regime: Measurement of fractional charge},}\ }\href {\doibase
  10.1126/science.267.5200.1010} {\bibfield  {journal} {\bibinfo  {journal}
  {Science}\ }\textbf {\bibinfo {volume} {267}},\ \bibinfo {pages} {1010--1012}
  (\bibinfo {year} {1995})}\BibitemShut {NoStop}%
\bibitem [{\citenamefont {Wen}(2002{\natexlab{a}})}]{WenPSG}%
  \BibitemOpen
  \bibfield  {author} {\bibinfo {author} {\bibfnamefont {Xiao-Gang}\
  \bibnamefont {Wen}},\ }\bibfield  {title} {\enquote {\bibinfo {title}
  {{Quantum orders and symmetric spin liquids}},}\ }\href {\doibase
  10.1103/PhysRevB.65.165113} {\bibfield  {journal} {\bibinfo  {journal} {Phys.
  Rev. B}\ }\textbf {\bibinfo {volume} {65}},\ \bibinfo {pages} {165113}
  (\bibinfo {year} {2002}{\natexlab{a}})}\BibitemShut {NoStop}%
\bibitem [{\citenamefont {Wen}(2002{\natexlab{b}})}]{Wen2002175}%
  \BibitemOpen
  \bibfield  {author} {\bibinfo {author} {\bibfnamefont {Xiao-Gang}\
  \bibnamefont {Wen}},\ }\bibfield  {title} {\enquote {\bibinfo {title}
  {Quantum order: a quantum entanglement of many particles},}\ }\href {\doibase
  https://doi.org/10.1016/S0375-9601(02)00808-3} {\bibfield  {journal}
  {\bibinfo  {journal} {Physics Letters A}\ }\textbf {\bibinfo {volume}
  {300}},\ \bibinfo {pages} {175 -- 181} (\bibinfo {year}
  {2002}{\natexlab{b}})}\BibitemShut {NoStop}%
\bibitem [{\citenamefont {Essin}\ and\ \citenamefont
  {Hermele}(2013)}]{PhysRevB.87.104406}%
  \BibitemOpen
  \bibfield  {author} {\bibinfo {author} {\bibfnamefont {Andrew~M.}\
  \bibnamefont {Essin}}\ and\ \bibinfo {author} {\bibfnamefont {Michael}\
  \bibnamefont {Hermele}},\ }\bibfield  {title} {\enquote {\bibinfo {title}
  {Classifying fractionalization: Symmetry classification of gapped
  ${\mathbb{z}}_{2}$ spin liquids in two dimensions},}\ }\href {\doibase
  10.1103/PhysRevB.87.104406} {\bibfield  {journal} {\bibinfo  {journal} {Phys.
  Rev. B}\ }\textbf {\bibinfo {volume} {87}},\ \bibinfo {pages} {104406}
  (\bibinfo {year} {2013})}\BibitemShut {NoStop}%
\bibitem [{\citenamefont {Essin}\ and\ \citenamefont
  {Hermele}(2014)}]{PhysRevB.90.121102}%
  \BibitemOpen
  \bibfield  {author} {\bibinfo {author} {\bibfnamefont {Andrew~M.}\
  \bibnamefont {Essin}}\ and\ \bibinfo {author} {\bibfnamefont {Michael}\
  \bibnamefont {Hermele}},\ }\bibfield  {title} {\enquote {\bibinfo {title}
  {Spectroscopic signatures of crystal momentum fractionalization},}\ }\href
  {\doibase 10.1103/PhysRevB.90.121102} {\bibfield  {journal} {\bibinfo
  {journal} {Phys. Rev. B}\ }\textbf {\bibinfo {volume} {90}},\ \bibinfo
  {pages} {121102} (\bibinfo {year} {2014})}\BibitemShut {NoStop}%
\bibitem [{\citenamefont {Banerjee}\ \emph {et~al.}(2008)\citenamefont
  {Banerjee}, \citenamefont {Isakov}, \citenamefont {Damle},\ and\
  \citenamefont {Kim}}]{PhysRevLett.100.047208}%
  \BibitemOpen
  \bibfield  {author} {\bibinfo {author} {\bibfnamefont {Argha}\ \bibnamefont
  {Banerjee}}, \bibinfo {author} {\bibfnamefont {Sergei~V.}\ \bibnamefont
  {Isakov}}, \bibinfo {author} {\bibfnamefont {Kedar}\ \bibnamefont {Damle}}, \
  and\ \bibinfo {author} {\bibfnamefont {Yong~Baek}\ \bibnamefont {Kim}},\
  }\bibfield  {title} {\enquote {\bibinfo {title} {{Unusual Liquid State of
  Hard-Core Bosons on the Pyrochlore Lattice}},}\ }\href {\doibase
  10.1103/PhysRevLett.100.047208} {\bibfield  {journal} {\bibinfo  {journal}
  {Phys. Rev. Lett.}\ }\textbf {\bibinfo {volume} {100}},\ \bibinfo {pages}
  {047208} (\bibinfo {year} {2008})}\BibitemShut {NoStop}%
\bibitem [{\citenamefont {Lv}\ \emph {et~al.}(2015)\citenamefont {Lv},
  \citenamefont {Chen}, \citenamefont {Deng},\ and\ \citenamefont
  {Meng}}]{PhysRevLett.115.037202}%
  \BibitemOpen
  \bibfield  {author} {\bibinfo {author} {\bibfnamefont {Jian-Ping}\
  \bibnamefont {Lv}}, \bibinfo {author} {\bibfnamefont {Gang}\ \bibnamefont
  {Chen}}, \bibinfo {author} {\bibfnamefont {Youjin}\ \bibnamefont {Deng}}, \
  and\ \bibinfo {author} {\bibfnamefont {Zi~Yang}\ \bibnamefont {Meng}},\
  }\bibfield  {title} {\enquote {\bibinfo {title} {{Coulomb Liquid Phases of
  Bosonic Cluster Mott Insulators on a Pyrochlore Lattice}},}\ }\href {\doibase
  10.1103/PhysRevLett.115.037202} {\bibfield  {journal} {\bibinfo  {journal}
  {Phys. Rev. Lett.}\ }\textbf {\bibinfo {volume} {115}},\ \bibinfo {pages}
  {037202} (\bibinfo {year} {2015})}\BibitemShut {NoStop}%
\bibitem [{\citenamefont {Kato}\ and\ \citenamefont
  {Onoda}(2015)}]{PhysRevLett.115.077202}%
  \BibitemOpen
  \bibfield  {author} {\bibinfo {author} {\bibfnamefont {Yasuyuki}\
  \bibnamefont {Kato}}\ and\ \bibinfo {author} {\bibfnamefont {Shigeki}\
  \bibnamefont {Onoda}},\ }\bibfield  {title} {\enquote {\bibinfo {title}
  {{Numerical Evidence of Quantum Melting of Spin Ice: Quantum-to-Classical
  Crossover}},}\ }\href {\doibase 10.1103/PhysRevLett.115.077202} {\bibfield
  {journal} {\bibinfo  {journal} {Phys. Rev. Lett.}\ }\textbf {\bibinfo
  {volume} {115}},\ \bibinfo {pages} {077202} (\bibinfo {year}
  {2015})}\BibitemShut {NoStop}%
\bibitem [{\citenamefont {Wen}(2002{\natexlab{c}})}]{PhysRevB.65.165113}%
  \BibitemOpen
  \bibfield  {author} {\bibinfo {author} {\bibfnamefont {Xiao-Gang}\
  \bibnamefont {Wen}},\ }\bibfield  {title} {\enquote {\bibinfo {title}
  {Quantum orders and symmetric spin liquids},}\ }\href {\doibase
  10.1103/PhysRevB.65.165113} {\bibfield  {journal} {\bibinfo  {journal} {Phys.
  Rev. B}\ }\textbf {\bibinfo {volume} {65}},\ \bibinfo {pages} {165113}
  (\bibinfo {year} {2002}{\natexlab{c}})}\BibitemShut {NoStop}%
\bibitem [{\citenamefont {Onoda}\ and\ \citenamefont
  {Tanaka}(2011)}]{PhysRevB.83.094411}%
  \BibitemOpen
  \bibfield  {author} {\bibinfo {author} {\bibfnamefont {Shigeki}\ \bibnamefont
  {Onoda}}\ and\ \bibinfo {author} {\bibfnamefont {Yoichi}\ \bibnamefont
  {Tanaka}},\ }\bibfield  {title} {\enquote {\bibinfo {title} {{Quantum
  fluctuations in the effective pseudospin-$\frac{1}{2}$ model for magnetic
  pyrochlore oxides}},}\ }\href {\doibase 10.1103/PhysRevB.83.094411}
  {\bibfield  {journal} {\bibinfo  {journal} {Phys. Rev. B}\ }\textbf {\bibinfo
  {volume} {83}},\ \bibinfo {pages} {094411} (\bibinfo {year}
  {2011})}\BibitemShut {NoStop}%
\bibitem [{\citenamefont {Li}\ \emph {et~al.}(2016)\citenamefont {Li},
  \citenamefont {Wang},\ and\ \citenamefont {Chen}}]{PhysRevB.94.201114}%
  \BibitemOpen
  \bibfield  {author} {\bibinfo {author} {\bibfnamefont {Yao-Dong}\
  \bibnamefont {Li}}, \bibinfo {author} {\bibfnamefont {Xiaoqun}\ \bibnamefont
  {Wang}}, \ and\ \bibinfo {author} {\bibfnamefont {Gang}\ \bibnamefont
  {Chen}},\ }\bibfield  {title} {\enquote {\bibinfo {title} {Hidden multipolar
  orders of dipole-octupole doublets on a triangular lattice},}\ }\href
  {\doibase 10.1103/PhysRevB.94.201114} {\bibfield  {journal} {\bibinfo
  {journal} {Phys. Rev. B}\ }\textbf {\bibinfo {volume} {94}},\ \bibinfo
  {pages} {201114} (\bibinfo {year} {2016})}\BibitemShut {NoStop}%
\end{thebibliography}%

\end{document}